\documentclass[a4paper]{article}
\usepackage[margin=25mm]{geometry}

\usepackage[version=4]{mhchem}
\usepackage{bm}
\usepackage{xifthen}
\usepackage{graphicx}
\usepackage{caption}
\usepackage{subcaption}
\graphicspath{{images/}}
\usepackage{float}
\usepackage[hidelinks]{hyperref}
\usepackage{amsfonts}
\usepackage{graphicx}
\usepackage{verbatim}
\makeatletter
\def\@seccntformat#1{\@ifundefined{#1@cntformat}%
   {\csname the#1\endcsname\quad}  
   {\csname #1@cntformat\endcsname}
}
\let\oldappendix\appendix 
\renewcommand\appendix{%
    \oldappendix
    \newcommand{\section@cntformat}{\appendixname~\thesection:~}
}
\makeatother

\usepackage{fancyhdr}
\usepackage[capitalize,nameinlink]{cleveref}[0.19]
\crefname{section}{section}{sections}
\crefname{subsection}{subsection}{subsections}
\Crefname{section}{Section}{Sections}
\Crefname{subsection}{Subsection}{Subsections}

\Crefname{figure}{Figure}{Figures}

\crefformat{equation}{\textup{#2(#1)#3}}
\crefrangeformat{equation}{\textup{#3(#1)#4--#5(#2)#6}}
\crefmultiformat{equation}{\textup{#2(#1)#3}}{ and \textup{#2(#1)#3}}
{, \textup{#2(#1)#3}}{, and \textup{#2(#1)#3}}
\crefrangemultiformat{equation}{\textup{#3(#1)#4--#5(#2)#6}}%
{ and \textup{#3(#1)#4--#5(#2)#6}}{, \textup{#3(#1)#4--#5(#2)#6}}{, and \textup{#3(#1)#4--#5(#2)#6}}

\Crefformat{equation}{#2Equation~\textup{(#1)}#3}
\Crefrangeformat{equation}{Equations~\textup{#3(#1)#4--#5(#2)#6}}
\Crefmultiformat{equation}{Equations~\textup{#2(#1)#3}}{ and \textup{#2(#1)#3}}
{, \textup{#2(#1)#3}}{, and \textup{#2(#1)#3}}
\Crefrangemultiformat{equation}{Equations~\textup{#3(#1)#4--#5(#2)#6}}%
{ and \textup{#3(#1)#4--#5(#2)#6}}{, \textup{#3(#1)#4--#5(#2)#6}}{, and \textup{#3(#1)#4--#5(#2)#6}}

\crefdefaultlabelformat{#2\textup{#1}#3}

\newcommand{\footremember}[2]{%
    \footnote{#2}
    \newcounter{#1}
    \setcounter{#1}{\value{footnote}}%
}
 
\title{Enzyme kinetics simulation at the scale of individual particles} 

\author{%
  Taylor Kearney\footremember{alley}{Monash University, Clayton, Victoria, Australia, Taylor.Kearney1@monash.edu}%
  \and Mark B. Flegg\footremember{trailer}{Monash University, Clayton, Victoria, Australia, Mark.Flegg@monash.edu}%
  }
\date{}

\providecommand{\keywords}[1]
{
  \small	
  \textbf{{Keywords---}} #1
}
\newcommand{\pos}[2]{\boldsymbol{#1}_{#2}}
\newcommand{\posbar}[2]{\bar{\boldsymbol{#1}}_{#2}}
\newcommand{\parentheses}[3]{\left#1#2\right#3}
\newcommand{\Eta}{\mathcal{H}}
\newcommand{\prob}[3][]{\ifthenelse{\isempty{#1}}{P\left(#2,#3\right)}{P\left(#1,#2,#3\right)}}
\newcommand{\probGiven}[4][]{\ifthenelse{\isempty{#1}}{P\parentheses{(}{#2,#3|#4}{)}}{P\parentheses{(}{#1,#2,#3|#4}{)}}}
\newcommand{\phiFunc}{\phi\left(\pos{\eta}{3},t\right)}
\newcommand{\phiFuncPrime}{\phi\left(\pos{\eta'}{3},t\right)}
\newcommand{\PhiFunc}{\Phi\left(\pos{\eta}{3},t\right)}
\newcommand{\gFunc}{g\left(\pos{\eta}{3},t\right)}

\begin{document}

\maketitle

\begin{abstract}
Enzyme-catalysed reactions involve two distinct timescales. There is a short timescale on which enzymes bind to substrate molecules to produce bound complexes, and a comparatively long timescale on which the complex is transformed into a product. The rate at which the substrate is converted into product is characteristically non-linear and is traditionally derived by applying singular perturbation theory to the system's governing equations. Central to this analysis is the assumption that complex formation is effectively instantaneous on the timescale over which significant substrate degradation occurs. This prevents accurate modelling of enzyme kinetics by many particle-based simulations of reaction-diffusion systems as they rely on proximity-based reaction conditions that do not correctly model the fast reactions associated with the complex on the long timescale. In this paper we derive a new proximity-based reaction condition that correctly incorporates the reactions that occur on the short timescale for a specific enzymatic system. We present proof of concept particle-based simulations and demonstrate that non-linear reaction rates typical of enzyme kinetics can be reproduced without needing to explicitly simulate reactions on the short timescale.
\end{abstract}


\section{Introduction}
Whole cell models promise to revolutionise systems biology. The capability to simulate the integrated function of every gene and molecule in a cell would assist clinicians in individualising therapy \cite{hamburg2010path,nielsen2017systems,jackson2015personalised} and enable computer-aided designs in synthetic biology \cite{whole_cell_synthetic_bio,10.3389/fbioe.2020.00942}. Their development would encourage the unification of our currently disconnected and heterogeneous biological datasets \cite{CARRERA2015719, 10.1093/nar/gks1108} and facilitate the discovery of emergent phenomena \cite{thornburg2022fundamental}. This worthy goal has been touted as a `grand challenge' for $21^{\text{st}}$ century systems biology \cite{TOMITA2001205} and will require extensive interdisciplinary collaboration if we are to be successful \cite{KARR201518}.\par
Early models attempted to describe the entirety of a cell's function using a single mathematical technique; namely ordinary differential equations (ODEs) \cite{Og_cell_model}. ODE models alone are not sufficient to describe all robust cellular behaviours. Over time, the disparate spatial and temporal scales of the involved phenomena inspired the development of hybrid models that are a conglomerate of many submodels that each target a specific biological module within the cell \cite{hartwell1999molecular, SZIGETI20188, KARR2012389}. Despite this progress, ODE models still abound and are typically used to describe the chemical reactions that dictate a cell's function \cite{chen2010classic}. Such models are underpinned by the presumption that the chemical species involved can be accurately represented as well-mixed, deterministic, time-varying continuous concentrations. However, in biological systems, molecules can occur in very low numbers and in highly localised distributions. For example, an entire cell may only contain a single molecule of mRNA for a particular gene \cite{copyNumbers}. When molecules become so sparsely distributed, it is impossible to select a neighbourhood about a point that contains sufficiently many molecules to define a meaningful concentration \cite{erban2009stochastic}. Moreover, cellular functions exhibit inherent stochasticity \cite{doi:10.1126/science.1147888,10.1371/journal.pcbi.1002010} owing to their origin in molecular interactions. In this regime, concentrations must be replaced by a collection of individual molecules undergoing reaction-diffusion processes. \par
Ideally, we explicitly model the molecular dynamics \cite{HOLLINGSWORTH20181129} that govern biochemical reactions, but such an approach results in simulations too computationally intensive for current computing hardware \cite{feig2019whole}. Failing this, we are forced to make simplifying assumptions about the involved physical processes in the hopes of obtaining a model that represents individual molecules, but abstains from explicit calculation of the intricate molecular interactions. Smoluchowski proposed such a model in $1917$ that describes the interaction of chemicals as diffusive point particles on a continuous domain, and has become one of the most widely accepted idealised models for reaction-diffusion systems of molecules \cite{smoluchowski1917versuch}.  Characteristic of this approach is the presumption that the system is sparse and that the relevant molecules can be treated as individuals that undergo isotropic diffusion as a result of their collisions with implicit solvent molecules. Bimolecular reactions are modelled by imposing that two molecules undergo a reaction if they become separated by less than a predefined distance $\sigma$. When a reaction occurs, the reactant molecules are removed from the system and replaced with a single molecule of the product of the reaction. We note for the sake of completeness that in Smoluchowski based frameworks unimolecular reactions are assumed to occur instantaneously and can be modelled as Poissonian processes that are independent of molecular diffusion \cite{Andrews_2004,egfrdAllDim}. Higher order reactions can also be incorporated by way of an extension developed by Flegg \cite{flegg2016smoluchowski} that we will review briefly in Section~\ref{sec:Generalized_Smoluchowski_theory}.\par
Smoluchowski's original reaction condition has been criticised for neglecting several important physical mechanisms that can influence reaction rates. These critiques have inspired the development of many derivative models that attempt to account for additional mechanisms including: activation energies \cite{collins1949diffusion}, intermolecular forces \cite{Debye_1942, KRAMERS1940284} and hydrodynamic effects \cite{HONIG197197,Wolynes&Deutch}. Despite this apparent diversity and in some cases the introduction of additional reaction parameters - see for instance models by Collins and Kimball \cite{collins1949diffusion}, or Doi \cite{Doi_1976,erban2009stochastic} - all current derivatives of Smoluchowski's reaction condition still describe bimolecular reactions as a proximity-dependent interaction between diffusing point particles. In each case, the definition of $\sigma$ can be altered to include extra information relating to the additional physical mechanisms considered, but its role as a parameter that summarises the molecular interaction remains unchanged \cite{flegg2016smoluchowski}. This underlying commonality is a testament to the robustness of Smoluchowski's original theory, and it serves as the foundation for many prominent software packages for particle-based simulation of reaction-diffusion systems, including: MCell \cite{MCell1,MCell2}, Smoldyn \cite{Andrews_2004}, Green’s function reaction dynamics (GFRD) \cite{gfrd1,gfrd2}, ReaDDy \cite{schoneberg2013readdy} and enhanced Green's function reaction dynamics (eGFRD) \cite{egfrd1,egfrd2}. All of these packages are capable of accurately simulating elementary unimolecular and bimolecular reactions in isolation. Thus, it seems reasonable to conclude that the same software can accurately simulate reaction networks composed of multiple elementary reactions, but for many biochemical networks this approach overlooks a fundamental assumption of Smoluchowski's reaction condition.\par 
Let us apply Smolcuhowski's model to a reaction between two chemical species $A$ and $B$ which produces a product $C$. We assume that initially the molecules of $A$ and $B$ are distributed uniformly at random within a volume $V$. The central result of Smoluchowski's theory states that two molecules (originally modelled as hard spheres) diffusing in a sufficiently large volume $V$, will - after an initial transient ($t_s$) - come into contact (at distance $\sigma$) at a constant rate per unit time $K$ given by
\begin{equation}\label{eq:smol_steady_rate_constant}
    K = \frac{k}{V} = \frac{4\pi \sigma \hat{D}_2}{V}.
\end{equation}
Here $\hat{D}_2 = D_A + D_B$ is the relative diffusion coefficient, and $D_A$ and $D_B$ are the diffusion coefficients associated with a molecule of $A$ and $B$ respectively. Equation (\ref{eq:smol_steady_rate_constant}) allows us to select $\sigma$ so that the reaction rate of our model matches the reaction rate of our bimolecular reaction. Crucially, this relation is only valid once the distribution of $B$ about $A$ molecules (and vice versa) reaches a steady state, which usually happens very quickly; within $10$ns for a typical system \cite{comprehensiveChemicalKinetics}. During the transient $t_s$ before the steady state is established, the reaction rate is artificially inflated as any molecules of $A$ and $B$ that are initialised within a distance $\sigma$ of each other undergo an instantaneous reaction and others in the neighbourhood of $\sigma$ undergo an inflated reaction rate temporarily. Due to this, the validity of Smoluchowski's reaction condition hinges on the assumption that a negligible number of reactions occur during $t_s$. For this to be true, we require that the expected separation of reactant pairs is large in comparison to $\sigma$ upon initialisation. In other words, the concentrations of $A$ and $B$ need to be sufficiently small, or the reactants must have a sufficiently low affinity for one another so that the associated reaction radius $\sigma$ is small.\par 
These requirements quickly become problematic when examining biochemical networks, since their action is often facilitated by biological catalysts called enzymes. Enzymes bind selectively to compounds known as substrates to enable essential biochemical reactions. They are fundamental to life and play a critical role in metabolic processes, cell regulation and signal transduction \cite{murray2003mathematical,alberts2017molecular}. Enzymatic systems are typically characterised by two distinct timescales. There is a short timescale $t_c$ on which enzymes bind to substrate molecules to form bound complexes, and a comparatively long timescale $t_p$ on which molecules of the complex are converted to products, degrading the substrate molecule and freeing the bound enzyme in the process. The classical analysis of these systems is based on the pioneering work by Michaelis and Menten who derived the degradation rate of the substrate for a prototypical enzymatic system \cite{michaelis1913kinetik}. Although Michaelis and Menten did not employ the theory themselves, it has long been known that the characteristically non-linear degradation rate of the substrate on the long timescale can be obtained by applying singular perturbation theory to the system of ODEs that governs the reaction \cite{the_quasi-steady-state_assumption}. Central to the application of the theory is the assumption that the timescale $t_c$ is effectively instantaneous in comparison to $t_p$. In essence, it is assumed that free enzymes bind so quickly to any available substrate molecules that the amount of bound complex is always in an instantaneous steady state with the amount of substrate; a so called quasi or pseudo-steady-state.\par 
The pseudo-steady-state assumption poses a serious problem for Smoluchowski's model of bimolecular reactions. In order for the model to be valid, we require that relatively few reactions occur during the initial transient. Conversely, if we are to mimic the results of the singular perturbation analysis we require the reactions between the enzyme and substrate molecules to be effectively instantaneous. If we attempt to rectify this by explicitly simulating the dynamics on the fast timescale, then we would have to wait a prohibitively long time before the substrate concentration had degraded appreciably. The issue is further exacerbated because reactions involving enzymes are reversible, meaning that the enzyme and substrate are free to unbind or disassociate from each other before the substrate is converted to a product. Through this mechanism, substrate and enzyme molecules are reintroduced into the system at positions uncontrolled by the modeller. Several initialisation methods exist that allow us to avoid artificial geminate recombination events, which arise when the reactants are initialised too close together and so immediately react \cite{Andrews_2004, lipkova2011analysis}. However, these methods consider the newly created pair of molecules in isolation and do nothing to account for the artificial reactions that can occur if the molecules are placed too close to other reactants in the system. The related Collins Kimball model avoids the geminate recombination problem by altering the Smoluchoswki reaction condition so that reactions only occur with a fixed probability once reactants are deemed close enough \cite{collins1949diffusion, diffusion_with_back_reaction}. This reduces the time it takes the system to reach a steady state following initialisation, but the problematic transient is not removed entirely, and the fundamental issue remains the same.\par
Motivated by these issues, we propose a modification to Smoluchowski's original reaction condition that allows us to reproduce non-linear reaction rates in trimolecular systems that are reminiscent of those in observed in Michaelis-Menten kinetics. Our new reaction condition is informed by the singular perturbation analysis of a trimolecular enzymatic system and incorporates the influence of the complex formation without requiring events on $t_c$ to be explicitly modelled. This completely circumvents all the issues posed by the disparate timescales, and enables us to directly embed the results of singular perturbation theory within Smoluchowski's framework. The generalised reaction conditions can be easily implemented in current software packages, and we demonstrate this by presenting proof of concept simulations.


\section{\label{sec:Generalized_Smoluchowski_theory}Generalised Smoluchowski theory}
 Smoluchowski only considered bimolecular reactions, but his work can be generalised to incorporate reactions of any order, as demonstrated by Flegg \cite{flegg2016smoluchowski}. This extension is fundamental to our results in this paper, so we provide a brief summary, although for more details readers are directed to the paper by Flegg.  \par
Consider a system of $N$ distinct molecules that are initially well-mixed (distributed uniformly at random) within a domain $\Omega$ of volume $V$, where $V$ is finite but very large. In addition, let $D_i$ and $\mathbf{x}_i$ denote the respective diffusion constant and the $3$-dimensional position of the i-th molecule. Since our reaction conditions will be functions of the molecules' relative proximity, it is convenient to transform the coordinate system into diffusive Jacobi coordinates or separation coordinates defined by
\begin{align}
    \pos{\eta}{1} &= \posbar{x}{N} \label{eq:eta1_def}\\
    \pos{\eta}{i} &= \pos{x}{i} - \posbar{x}{i-1}, \quad i = 2,3,...,N,  \quad \text{where,}\label{eq:etai_def}\\
    \posbar{x}{i} &= \frac{\sum^i_{j=1} \pos{x}{j}D_j^{-1}}{\sum^i_{k=1}D_k^{-1}},\label{eq:x_bari_def}
\end{align}
is the `centre of diffusion' of the first $i$ molecules (analogous to the centre of mass except that the positions are weighted by their inverse diffusion coefficients rather than their masses).
In our new coordinate system $\pos{\eta}{1}$ is the centre of diffusion of the $N$ molecules and $\pos{\eta}{i}$ for $i > 1$  describes the separation between $\pos{x}{i}$ and the centre of diffusion, $\posbar{x}{i-1}$, of the previous $i-1$ molecules. The state vector $\pos{\eta}{i}$ can be shown to undergo independent linear diffusion with diffusion constant
\begin{equation}\label{eq:eta_i_diff_coeff}
    \hat{D}_i = \begin{cases}
                    \bar{D}_N \quad \text{ when } i = 1,\\
                    D_i + \bar{D}_{i-1} \quad \text{ when } i>1,
                \end{cases}
\end{equation}
where $\bar{D}_j$ is the diffusion constant associated with $\posbar{x}{j}$,
\begin{equation}\label{eq:x_bar_diff_coeff}
    \bar{D}_j = \frac{1}{\sum_{i=1}^j D_i^{-1}}.
\end{equation}\par
Under this specific transformation, the molecules, and by extension the state vectors, diffuse independently. This framework is useful to work in since the physical constraint that `translations of the whole system should not cause a reaction' can be simply translated as `reaction conditions must be independent of $\pos{\eta}{1}$'. As a result, for a well-mixed system, the joint probability density, $\prob{\pos{\eta}{}}{t}$, to find the reactants in an unreacted state is also independent of $\pos{\eta}{1}$ and is given by the diffusion equation
\begin{equation}\label{eq:general_prob_density_PDE}
    \frac{\partial \prob{\pos{\eta}{}}{t}}{\partial t} = \left[\sum_{i=2}^N \hat{D}_i\hat{\nabla}^2_i \right]\prob{\pos{\eta}{}}{t},
\end{equation}
where $\boldsymbol{\eta} =  \left\{\pos{\eta}{2},\pos{\eta}{3},...,\pos{\eta}{N}\right\}$ describes the state of the system and $\hat{\nabla}^2_i$ denotes the Laplacian with respect to the coordinates of $\pos{\eta}{i}$. Initially, no reactions have occurred, and the probability density can be found by normalisation,
\begin{equation}\label{eq:general_prob_density_IC}
    P\left(\boldsymbol{\eta},0\right) = P_\infty= \frac{1}{V^{N-1}},
\end{equation}
noting that the power here is $N-1$ since the state does not depend on the coordinates of $\pos{\eta}{1}$ and so $P$ is the probability density to find the other coordinates only.
The inner boundary corresponds to the reaction condition and defines a region, $\Omega_R$, upon whose boundary, $\partial \Omega_R$, the system is absorbed and reacts,
\begin{equation}\label{eq:general_prob_density_IB}
    P\left(\boldsymbol{\eta} \in \partial \Omega_R,t\right) = 0.
\end{equation}
Sufficiently far from the origin we expect $P$ to be unperturbed by the absorption of states on the inner boundary, and we require
\begin{equation}\label{eq:general_prob_density_OB}
    \lim_{\boldsymbol{\eta} \to \infty} P\left(\boldsymbol{\eta},t\right) = P_\infty= \frac{1}{V^{N-1}}.
\end{equation}\par
To recover Smoluchowski's original result we restrict ourselves to $N=2$ and adopt the inner boundary
\begin{equation}\label{eq:smol_bimolecular_condition}
    \Omega_R = \left\{\pos{\eta}{} = \pos{\eta}{2} : r_2 \le \sigma\right\}, \quad \text{where } r_2 = ||\pos{\eta}{2}||.    
\end{equation}
Assuming the system is initially well-mixed and that $\sigma$ is comparatively small, we can solve Equation (\ref{eq:general_prob_density_PDE}) via Laplace transform which yields the radially symmetric solution \cite{comprehensiveChemicalKinetics}
\begin{equation}\label{eq:smol_prob_density}
    P\left(r_2,t\right) = \frac{1}{V}\left[1-\frac{\sigma}{r_2}\text{erfc}\left(\frac{r_2-\sigma}{\sqrt{4\hat{D}_2t}}\right)\right].
\end{equation}
The corresponding reaction rate is given by the total flux of the probability density over the inner boundary,
\begin{equation}\label{eq:smol_time_dependent_rate}
    K = \frac{4\pi \sigma \hat{D}_2}{V}\left(1+\frac{\sigma}{\sqrt{4\hat{D}_2t}}\right).
\end{equation}\par
We see that in general the probability density and hence Smoluchowski's reaction rate, is time dependent. The time dependent rate in Equation (\ref{eq:smol_time_dependent_rate}) is only well approximated by the more commonly stated steady-state rate in Equation (\ref{eq:smol_steady_rate_constant}) once the probability density has converged sufficiently to steady-state distribution 
\begin{equation}
    \lim_{t\to \infty}P\left(r_2,t\right) = \frac{1}{V}\left(1-\frac{\sigma}{r_2}\right).
\end{equation}
Before the system reaches this steady state the reaction rate can be arbitrarily high, diverging to infinity as we approach $t=0$. This transient behaviour is an artefact of the initial condition, where two reactants have a non-zero probability of being initialised such that $r_2 \le \sigma$ (or even an elevated chance of being initialised with $r_2\sim \sigma$). Any such pair will undergo an instantaneous (or increased rate of) reaction that is not diffusion limited, i.e. the reaction rate is not determined by reactant diffusion. If the expected separation between reactants is sufficiently large in comparison to $\sigma$ then reactions that occur during the transient can be neglected without incurring significant errors. This is the antithesis of the conditions required for the pseudo-steady-state assumption to be valid. Instead, we expect that a significant number of reactions will occur during this transient, which would cause the reaction rate to exceed what is predicted by Equation (\ref{eq:smol_steady_rate_constant}).


\section{\label{sec:generalised_reaction_conditions}Generalised reaction conditions}
To illustrate the aforementioned issues with Smoluchowski's reaction condition and our proposed solution, we consider a network that consists of three reactions and involves three chemical species $A$, $B$ and $C$. In the first reaction, molecules of $A$ and $B$ are able to bind to form a chemical complex that we will denote $X$. Once formed, a molecule of $X$ is able to disassociate into its constituent $A$ and $B$ molecules, or it can react with a molecule of $C$, converting the complex to some product $P$. The production of $P$ does not consume $C$ molecules and instead frees them to participate in other reactions. Our reaction network can be summarised succinctly in chemical shorthand as
\begin{equation}\label{eq:Example_reaction_network}
    \ce{A + B <=>[k_1][k_{-1}] X} \quad \text{and} \quad \ce{X + C ->[k_2] P + C},
\end{equation}
where $k_1$ and $k_2$ are the second-order rate constants that govern the formation rate of $X$ and $P$ respectively, while $k_{-1}$ is the first-order rate constant that controls the rate of dissociation of $X$ \cite{CORNISHBOWDEN19791}.\par
The Law of Mass Action states that the rate of an elementary reaction is proportional to the product of the reactant concentrations, with the constant of proportionality being the rate constant associated with the reaction. Applying this to Reaction (\ref{eq:Example_reaction_network}) yields a system of five ODEs that can be reduced to just two equations,
\begin{align}
    \frac{da}{dt} &= -k_1ab + k_{-1}x\label{eq:Example_mass_action_reduced1}, \quad \text{and}\\
    \frac{dx}{dt} &= k_1ab - k_{-1}x - k_2xc\label{eq:Example_mass_action_reduced2},
\end{align}
where the lower case letters denote the concentrations of the corresponding chemical species.
The traditional approach - more rigorously justified by Briggs and Haladane \cite{A_note_on_the_kinetics_of_enzyme_action} than Michaelis and Menten \cite{michaelis1913kinetik} - now proceeds by applying the pseudo-steady-state approximation to Equations (\ref{eq:Example_mass_action_reduced1}) and (\ref{eq:Example_mass_action_reduced2}). The approximation is often stated verbatim, but it is more instructive to view it as the result of singular perturbation theory. To obtain non-linear kinetics we require the complex concentration to be in a pseudo-steady-state on the long timescale. For this to occur the complex needs to be short-lived so will we consider a situation where it dissociates much more quickly than it is formed. In addition, we assume the concentration of $C$ is much larger than that of $A$, or $B$, so that reactions between $X$ and $C$ occur much faster than those between $A$ and $B$. That is, we assume that both $k_{-1}$ and the initial concentration of $C$, $c_0$, are $O\left(\frac{1}{\varepsilon}\right)$ for a sufficiently small positive dimensionless parameter $\varepsilon$ such that,
\begin{equation}
    k_{-1} = \frac{\bar{k}_{-1}}{\varepsilon} \quad \text{and} \quad c_0 = \frac{\bar{c}_0}{\varepsilon},
\end{equation}
where $\bar{k}_{-1}$ and $\bar{c}_0$ are both $O(1)$.\par
To proceed we define the dimensionless variables
\begin{equation}\label{eq:Example_dimensionless_vars}
    \bar{a} = \frac{a}{a_0}, \quad \bar{b} = \frac{b}{b_0}, \quad \bar{c} = \frac{c}{c_0},\quad \bar{x} = \frac{x}{a_0}, \quad \text{and} \quad T = k_1b_0t,
\end{equation}
where $a_0$ and $b_0$ are the initial concentrations of $A$ and $B$ respectively. Applying our change of variables to Equations (\ref{eq:Example_mass_action_reduced1}) and (\ref{eq:Example_mass_action_reduced2}) gives
\begin{align}
    \varepsilon\frac{d\bar{a}}{dT} &= -\varepsilon\bar{a}\bar{b} + \mu\bar{x},\label{eq:dimesionless_system1}\quad \text{and}\\
    \varepsilon\frac{d\bar{x}}{dT} &=  \varepsilon\bar{a}\bar{b} - \mu\bar{x} - \nu\bar{x}\bar{c},\label{eq:dimesionless_system2}
\end{align}
where we have defined the dimensionless parameters
\begin{equation}\label{eq:dimensionless_params}
    \mu = \frac{\bar{k}_{-1}}{k_1b_0}, \quad \text{and} \quad \nu = \frac{k_2\bar{c}_0}{k_1b_0}.
\end{equation}
From Equation (\ref{eq:dimesionless_system1}) we find that $\bar{x}$ is $O(\varepsilon)$ in which case Equation (\ref{eq:dimesionless_system2}) yields
\begin{equation}\label{eq:zeroth_order_soln_x}
    \bar{x}\left(T\right) = \frac{\varepsilon\bar{a}\bar{b}}{\mu+\nu\bar{c}} + O\left(\varepsilon^2\right).
\end{equation}
Substituting Equation (\ref{eq:zeroth_order_soln_x}) into Equation (\ref{eq:dimesionless_system1}) and redimensionalising gives to leading order in $\varepsilon$,
\begin{equation}\label{eq:MM_degradation_rate_a}
    \frac{da}{dt} = -\frac{k_1 abc}{\Gamma + c},
\end{equation}
where
\begin{equation}\label{eq:Michaelis_constant}
    \Gamma = \frac{k_{-1}}{k_2},
\end{equation}
is analogous to the Michaelis constant for Reaction (\ref{eq:Example_reaction_network}).\par
The preceding analysis shows that on the long timescale Reaction (\ref{eq:Example_reaction_network}) can be reduced to a single trimolecular reaction
\begin{equation}\label{eq:tri_molecular_reaction}
    \ce{A + B + C ->[k_3\parentheses{(}{c}{)}] P + C},
\end{equation}
where $k_3$ is the third-order rate constant
\begin{equation}\label{eq:trimolecular_rate_constant}
    k_3\parentheses{(}{c}{)} = \frac{k_1}{\Gamma + c}.
\end{equation}
This reduction is valid so long as $\varepsilon$ is sufficiently small. This requires that the rate at which a single molecule of $X$ forms is slow when compared to the rate at which it dissociates and/or the rate at which it is converted into the product. In our analysis we assumed that the reaction between $X$ and $C$ was fast due to $c_0$ being very large in comparison to $a_0$ and $b_0$, however we could have assumed instead that $k_2$ was much larger than $k_1$ and arrived at the same result.\par
\begin{figure}
\centering
\begin{subfigure}[b]{0.75\textwidth}
   \includegraphics[width=1\linewidth]{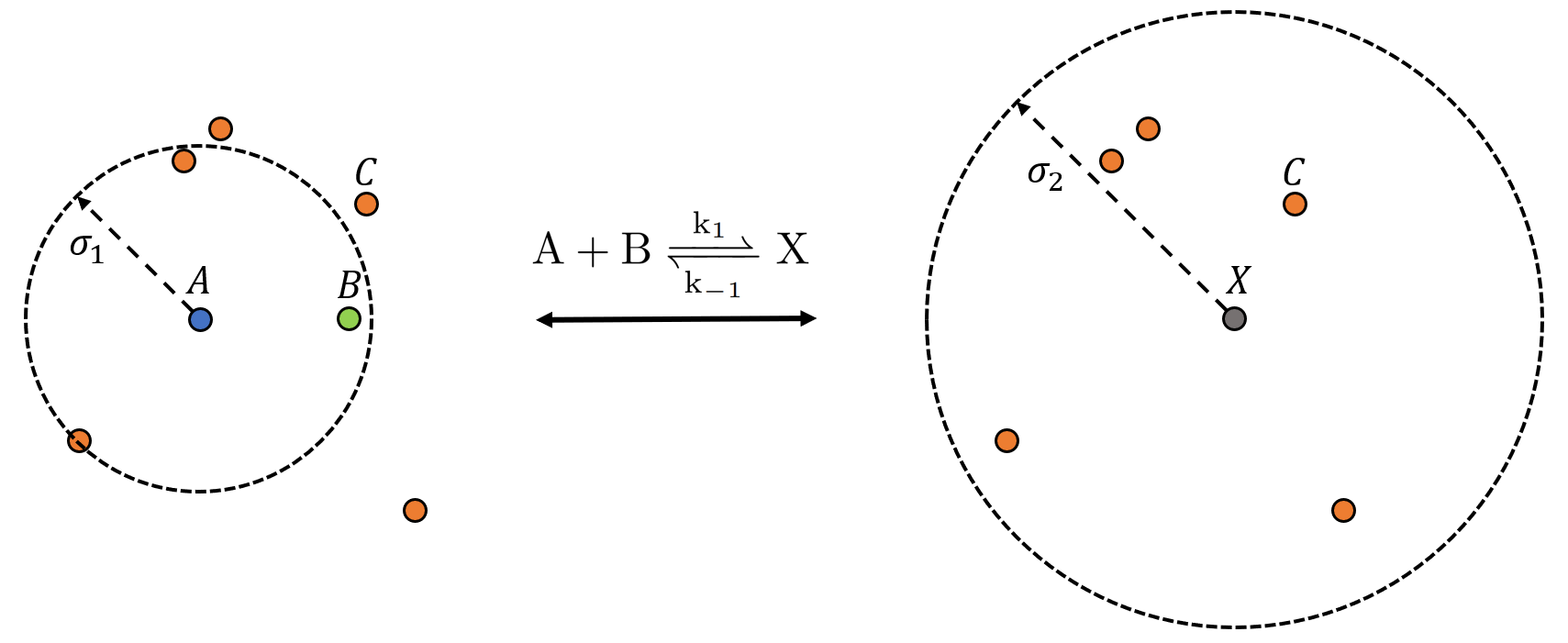}
   \caption{}
   \label{fig:smol_sim} 
\end{subfigure}
\begin{subfigure}[b]{0.75\textwidth}
   \includegraphics[width=1\linewidth]{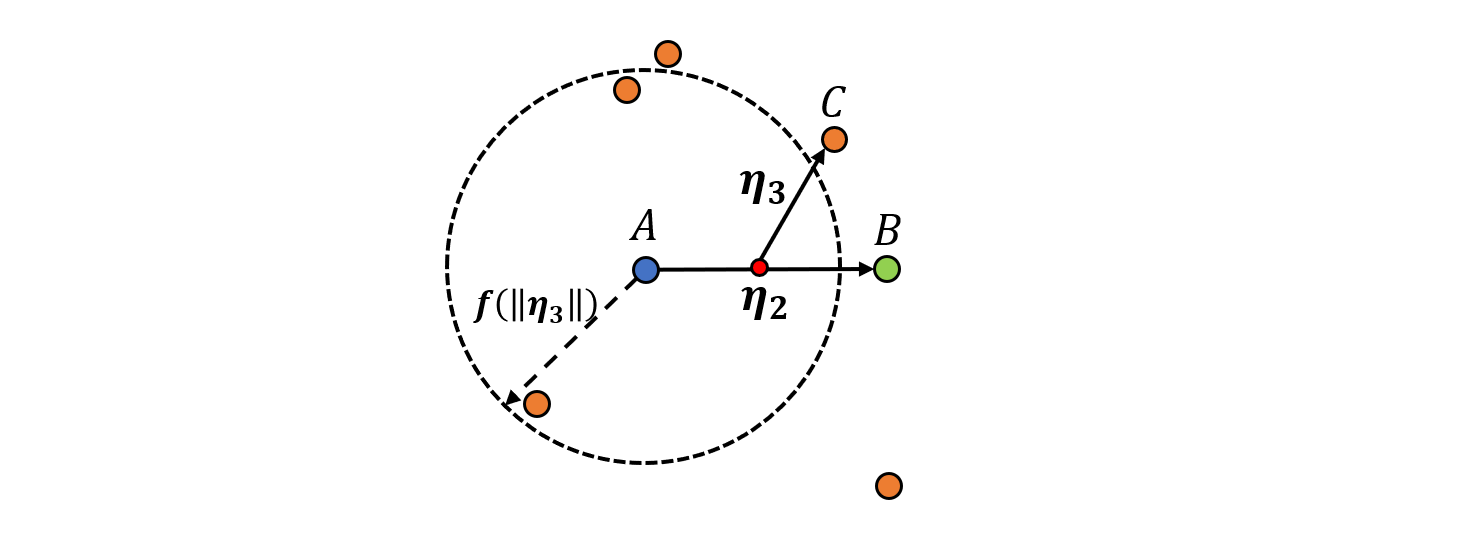}
   \caption{}
   \label{fig:tri_sim}
\end{subfigure}
\caption{(a) A particle-based simulation of Reaction (\ref{eq:Example_reaction_network}) that utilises the reaction condition given in Equation (\ref{eq:smol_bimolecular_condition}) to model the bimolecular reactions $\ce{A + B <=>[k_1][k_{-1}] X}$ and $\ce{X + C ->[k_2] P + C}$. On the left, a pair of molecules $A$ and $B$ - depicted as a blue and green point respectively - are simulated as undergoing a bimolecular reaction to form a molecule of $X$ whenever their separation is less than the reaction radius $\sigma_1$, which is related to the rate constant $k_1$ via Equation (\ref{eq:smol_steady_rate_constant}). Once a molecule of $X$ has formed - depicted as a grey point on the right - it is initialised with a reaction radius $\sigma_2$ which is proportional to $k_2$, again in accordance with Equation (\ref{eq:smol_steady_rate_constant}). If the reaction between $X$ and $C$ is fast (compared to the rate at which $X$ is formed) because either $k_2$, or the concentration of $C$ is large, then the expected separation between $X$ and molecules of $C$ becomes small in comparison to $\sigma_2$ leading to a reaction rate in excess of that predicted by Equation (\ref{eq:smol_steady_rate_constant}). Due to this, Reaction (\ref{eq:Example_reaction_network}) cannot be accurately simulated using the reaction condition given in Equation (\ref{eq:smol_bimolecular_condition}). (b) A particle-based simulation of Reaction (\ref{eq:tri_molecular_reaction}) which degrades molecules of $A$ and $B$ (equivalently produces molecules of $P$) at the same rate as Reaction (\ref{eq:Example_reaction_network}). A pair of molecules $A$ and $B$ - depicted as a blue and green point respectively - are simulated as undergoing a trimolecular reaction with a molecule of $C$ - shown by the orange points - if their separation $||\pos{\eta}{2}||$ is less than the distance $f\left(||\pos{\eta}{3}||\right)$. Unlike the reaction boundaries in part (a) of this figure, $f\left(||\pos{\eta}{3}||\right)$ is not static and is instead a function of separation $\pos{\eta}{3}$ between the centre of diffusion of $A$ and $B$ - shown by the small red point - and the closest molecule of $C$ to this point. The function $f$ is chosen so that the simulated reaction rate matches the non-linear rate of Reaction (\ref{eq:tri_molecular_reaction}) (see Equation (\ref{eq:generalised_MM_reaction_boundary}) for this particular form of $f$). By enabling the simulation of Reaction (\ref{eq:tri_molecular_reaction}) this method accurately reproduces the kinetics of Reaction (\ref{eq:Example_reaction_network}).}
\label{fig:difference_to_Smoluchowski}
\end{figure}

We have shown that under the pseudo-steady-state approximation Reaction (\ref{eq:Example_reaction_network}) is equivalent to Reaction (\ref{eq:tri_molecular_reaction}) on the long timescale. The disparate timescales required to make this approximation valid, also preclude accurate particle-based simulation of Reaction (\ref{eq:Example_reaction_network}) by any methods that rely on Smoluchowski's bimolecular reaction condition given in Equation (\ref{eq:smol_bimolecular_condition}). For instance, the fast bimolecular reaction between $X$ and $C$ cannot be accurately simulated on the long timescale using this condition as shown in Fig.~\ref{fig:smol_sim}. To avoid this issue, we seek instead construct a particle-based simulation of Reaction (\ref{eq:tri_molecular_reaction}), which requires the development of a new trimolecular reaction condition that reproduces the non-linear reaction rate in Equation (\ref{eq:trimolecular_rate_constant}) as shown in Fig.~\ref{fig:tri_sim}.\par

Suppose we alter our system slightly so that it now contains $N_C = cV$ molecules of $C$ - but still just a single molecule each of $A$ and $B$ - where, as before, $c$ is a well-mixed concentration of $C$ molecules. Since there are now multiple molecules of $C$ we must consider $N_C$ distinct states; one for each combination of $A$, $B$ and $C$ molecules. We recall that $\pos{\eta}{2}$ and $\pos{\eta}{3}$ diffuse independently so that, $P\parentheses{(}{\boldsymbol{\eta},t}{)} = P_2\parentheses{(}{\pos{\eta}{2},t}{)}P_3\parentheses{(}{\pos{\eta}{3},t}{)}$, and
\begin{align}
    \parentheses{[}{\frac{\partial}{\partial t} - \hat{D}_2\hat{\nabla}^2_2}{]}P_2\parentheses{(}{\pos{\eta}{2},t}{)} &\equiv\mathcal{L}_2P_2\parentheses{(}{\pos{\eta}{2},t}{)} = 0, \quad \text{and} \label{eq:diffusion_eta2}\\
    \parentheses{[}{\frac{\partial}{\partial t} - \hat{D}_3\hat{\nabla}^2_3}{]}P_3\parentheses{(}{\pos{\eta}{3},t}{)} &\equiv\mathcal{L}_3P_3\parentheses{(}{\pos{\eta}{3},t}{)} = 0\label{eq:diffusion_eta3},
\end{align}
where $\mathcal{L}_2$ and $\mathcal{L}_3$ are diffusion operators on the $3$-dimensional spaces defined by $\pos{\eta}{2}$ and $\pos{\eta}{3}$ respectively. Since there is just one molecule of $B$, all states (where each state consists of one $A$, one $B$ and one $C$ molecule) lie on manifolds of constant $\pos{\eta}{2}$, whilst the specific instance of $\pos{\eta}{2}$ also diffuses according to Equation (\ref{eq:diffusion_eta2}). On the manifold, states diffuse independently in the $\pos{\eta}{3}$ space in accordance with Equation (\ref{eq:diffusion_eta3}). The first state incident on the inner boundary $\partial \Omega$ will cause a reaction, and therefore we wish to know the dynamics of the state with the minimum magnitude $||\pos{\eta}{3}||$.\par

We need to understand the well-mixed steady state of this system. In the well-mixed state, we can assume that the $\pos{\eta}{3}$ coordinates of the $N_C$ states are uniformly and independently distributed in $\Omega$. Consider now a single particular molecule of $C$ and let $\pos{\Eta}{3}$ denote the event that this particular molecule is associated with the minimum $||\pos{\eta}{3}||$ when compared with any other $C$ molecule in the system. Moreover, let the probability distribution function, $\probGiven[]{\pos{\eta}{3}}{t}{\pos{\Eta}{3}}$, denote the probability density for finding this molecule of $C$ at $\pos{\eta}{3}$ at time $t$, given the fact that it has the smallest $||\pos{\eta}{3}||$ of any molecule of $C$. Since the molecules are well-mixed, the probability that an arbitrary $C$ molecule has the minimum $||\pos{\eta}{3}||$ is
\begin{equation}\label{eq:prob_eta3_given_Eta3}
    \prob[]{\pos{\Eta}{3}}{t} = \frac{1}{N_C}.
\end{equation}
The probability of $\pos{\Eta}{3}$, that a particular molecule of $C$ is closest to the origin, given it has a known $\pos{\eta}{3}$, is equal to the probability that all the other $N_C-1$ molecules of $C$ lie outside a sphere, $V_3$, of radius $r_3 = ||\pos{\eta}{3}||$ centred on the origin, i.e.
\begin{equation}
    \probGiven[]{\pos{\Eta}{3}}{t}{\pos{\eta}{3}} = \left[1 - \int_{V_3}\prob[]{\pos{\eta'}{3}}{t}dV_3'\right]^{N_C-1},
\end{equation}
where $dV_3'$ is an elemental volume for coordinates of $\pos{\eta'}{3}$. Bayes Theorem then yields
\begin{equation}\label{eq:bayes_theorem_sub}
     \probGiven[]{\pos{\eta}{3}}{t}{\pos{\Eta}{3}} = N_C\prob[]{\pos{\eta}{3}}{t}\left[1-\int_{V_3}\prob[]{\pos{\eta'}{3}}{t}dV_3'\right]^{N_C-1}.
\end{equation}
\par
The system is very large and in the limit that $V$ - and hence $N_C$ - tends to infinity, $\prob[]{\pos{\Eta}{3}}{t}$ goes to zero in accordance with Equation (\ref{eq:prob_eta3_given_Eta3}). To ensure we take the appropriate limit in Equation (\ref{eq:bayes_theorem_sub}) we define the scaled probability distributions
\begin{align}
    \PhiFunc &= \frac{\probGiven[]{\pos{\eta}{3}}{t}{\pos{\Eta}{3}}}{c}, \quad \text{and} \label{eq:Phi_definition}\\
    \phiFunc &= \frac{N_C\prob[]{\pos{\eta}{3}}{t}}{c} \label{eq:phi_definition},
\end{align}
which when substituted into Equation (\ref{eq:bayes_theorem_sub}) give,
\begin{equation}
    \PhiFunc = \phiFunc\left[1-\frac{c}{N_C}\int_{V_3}\phiFuncPrime dV_3'\right]^{N_C-1}.
\end{equation}
Taking the limit $N_C \rightarrow \infty$ we obtain,
\begin{equation}\label{eq:Phi_definition_in_limit}
    \PhiFunc = \phiFunc \text{exp}\left(-c\int_{V_3}\phiFuncPrime dV_3'\right).
\end{equation}
That is, using Equations (\ref{eq:diffusion_eta2}) and (\ref{eq:diffusion_eta3}), $\PhiFunc$ evolves according the diffusion-advection equation
\begin{equation}\label{eq:Phi_evolution}
    \frac{\partial \PhiFunc}{\partial t} = \hat{D}_3\hat{\nabla}^2_3\PhiFunc + \hat{D}_3\hat{\nabla}_3 \cdot \parentheses{(}{4\pi r_3^2 c\phi\PhiFunc\hat{\pos{r}{3}}}{)},
\end{equation}
where $\hat{\pos{r}{3}}$ is the unit outward facing normal vector of a sphere of radius $r_3$; a derivation of this result can be found in Appendix \ref{sec:AppendixA}. We note here that the isotropic linear diffusion term describes the independent Brownian motion of $\pos{\eta}{2}$ and $\pos{\eta}{3}$ whilst advection towards $\pos{\eta}{3}=\mathbf{0}$ represents the flux of the likelihood that the $C$ molecule with the second-smallest $||\pos{\eta}{3}||$ value diffuses over the sphere of radius $r_3$ set by the current $C$ molecule.  

Typically unless a boundary-free steady state in $\mathbb{R}^6$ is sought (see Equation (\ref{eq:Phi_definition_in_limit})), this PDE is very difficult to solve. This is because of the intrinsic relationship between $\phi$ and $\Phi$. In our particular case however, we will be assuming that there is a very thin absorbing boundary which is long in $r_3 = ||\pos{\eta}{3}||$ but thin in $r_2 = ||\pos{\eta}{2}||$. We expect therefore that $\phi$ is equal to its well-mixed value of $\phi = 1$ with a small perturbation caused by undulations of the absorbing surface. As we will only concern ourselves with the leading order solution of this PDE with a thin absorbing boundary, using $\phi=1$ and Equation (\ref{eq:Phi_definition}) we arrive at the governing equation for $\probGiven[]{\pos{\eta}{3}}{t}{\pos{\Eta}{3}}$,
\begin{equation}\label{PDEDiffAdv}
    \frac{\partial \probGiven[]{\pos{\eta}{3}}{t}{\pos{\Eta}{3}}}{\partial t} =  \hat{D}_3\hat{\nabla}^2_3\probGiven[]{\pos{\eta}{3}}{t}{\pos{\Eta}{3}}+ \hat{D}_3\hat{\nabla}_3 \cdot \parentheses{(}{4\pi c r_3^2 \probGiven[]{\pos{\eta}{3}}{t}{\pos{\Eta}{3}}\hat{\pos{r}{3}}}{)}.
\end{equation}
As $\pos{\eta}{2}$ diffuses independently of $\pos{\eta}{3}$ the evolution of the joint probability density, $\probGiven[]{\boldsymbol{\eta}}{t}{\pos{\Eta}{3}} = \probGiven[\pos{\eta}{2}]{\pos{\eta}{3}}{t}{\pos{\Eta}{3}}$, for finding the separation of the state with the minimum value of $\pos{\eta}{3}$ is governed by
\begin{equation}\label{eq:governing_min_eta3_PDE}
    \frac{\partial \probGiven[]{\boldsymbol{\eta}}{t}{\pos{\Eta}{3}}}{\partial t} =  \left[\sum_{i=2}^3 \hat{D}_i\hat{\nabla}^2_i \right]\probGiven[]{\boldsymbol{\eta}}{t}{\pos{\Eta}{3}} + \hat{D}_3\hat{\nabla}_3 \cdot \parentheses{(}{4\pi c r_3^2 \probGiven[]{\boldsymbol{\eta}}{t}{\pos{\Eta}{3}}\hat{\pos{r}{3}}}{)}.
\end{equation}

To find the correct boundary conditions for the probability $\probGiven[]{\boldsymbol{\eta}}{t}{\pos{\Eta}{3}}$ we need to remind ourselves that this probability is zero (absorbing boundary condition) if the state $\boldsymbol{\eta}$ ever reaches a manifold on which a reaction condition is met. We note that if the condition for a reaction occurs on an absorbing boundary extending small distances $r_2$ and $r_3$ then the advection term in Equation (\ref{PDEDiffAdv}) becomes negligible as was the case for the trimolecular generalisation of Smoluchowski reaction condition presented by Flegg \cite{flegg2016smoluchowski} (although never directly addressed in that paper). That is, any small absorbing boundary will reach a pseudo-equilibrium that returns trimolecular mass action. Instead, in order to obtain non-linear reaction kinetics in the concentration of $C$ we propose a long thin boundary $\partial \Omega_{NL}$ where
\begin{equation}\label{eq:generalised_condition}
    \Omega_{NL} = \left\{(\pos{\eta}{2}, \pos{\eta}{3}) : r_2 < f(r_3)\right\}, \quad r_2 = ||\pos{\eta}{2}||,
\end{equation}
where $0<f(r_3)<\sigma$ is a monotonically decreasing function of $r_3$ and $\sigma$ is a small positive constant relative to the diffusion coefficients $\hat{D}_i$ and the characteristic scale of the domain of $f$. Since this boundary is thin, to leading order the normal to the boundary is $\mathbf{\hat{n}} = \pos{\eta}{2}$ and the steady-state solution to Equation (\ref{eq:governing_min_eta3_PDE}) with $\probGiven[]{\boldsymbol{\eta}}{t}{\pos{\Eta}{3}} = 0$ on $\partial \Omega_{NL}$ where $\Omega_{NL}$ is given by Equation (\ref{eq:generalised_condition}) is weakly dependent on $\pos{\eta}{3}$ compared to $\pos{\eta}{2}$. That is, the problem reduces approximately to solving for the steady state of the problem
\begin{equation}\label{PDEDiffAdv2}
    \frac{\partial \prob{\pos{\eta}{2}}{t}}{\partial t} =  \hat{D}_2\hat{\nabla}^2_2\prob{\pos{\eta}{2}}{t},\quad \text{with }  \prob{\pos{\eta}{2}}{t} = 0 \text{ when } r_2 = f(r_3),
\end{equation}
and the value of $\prob{\pos{\eta}{2}}{t} = Q(r_3)$ at infinity is given by the steady-state solution of Equation (\ref{PDEDiffAdv}) on $\mathbb{R}^3$ -- that is, 
\begin{equation}
    Q(r_3) = c \exp(-4\pi r_3^3 c/3).
\end{equation}
\par
Finding the flux in the $\mathbf{\hat{n}}$ direction over $\partial \Omega_{NL}$ and matching it to reaction rate $K$ is the same as solving the bimolecular Smoluchowski reaction boundary problem for the reaction rate at each $r_3$ where the reaction radius is $r_2 = f(r_3)$, call this $K_2(f(r_3)))$ where $K(\rho) = 4\pi\hat{D}_2 \rho$ is the well known Smoluchowski result, multiplying this rate by the probability of finding a reaction with a given $r_3$, $Q(r_3)$, and integrating over all spheres of radius $r_3$ to find the total flux. That is,
\begin{equation}
    K = \int^\infty_0 K_2\left(f(r_3)\right)Q\left(r_3\right)4\pi r_{3}^2 dr_3.  
\end{equation}
Making the substitution $\lambda = 4\pi r_3^3/3$ and recalling that $K\equiv K(c)$ here is a rate that depends on the concentration $c$ for each molecule of $A$ and $B$,
\begin{equation}\label{eq:laplace_transform_rate}
    \frac{K(c)}{4\pi c \hat{D}_2} = \int^{\infty}_0 F(\lambda) \exp(-\lambda c) d\lambda = \mathcal{L}(F(\lambda)).  
\end{equation}
where $\mathcal{L}$ is the Laplace transform and $F(4\pi r_3^3 /3) = f(r_3)$. For Reaction (\ref{eq:tri_molecular_reaction}) the reaction rate is, $K(c) = ck_3(c)$ where $k_3(c)$ is given by Equation (\ref{eq:trimolecular_rate_constant}). In this case it is easy therefore to take the inverse Laplace transform to find $F$ and therefore $f$ (the unknown function that we require to construct a reaction condition),
\begin{equation}
    \label{eq:generalised_MM_reaction_boundary}
    f(r_3) = \frac{k_1}{4\pi\hat{D}_2}\text{exp}\left(\frac{-4\pi\Gamma r_3^3}{3}\right).
\end{equation}
Finally, substituting this result into Equation (\ref{eq:generalised_condition}) we obtain a reaction boundary $\partial \Omega_{NL}$,
\begin{equation}\label{eq:generalised_MM_reaction_condition}
    \Omega_{NL} = \left\{\pos{\eta}{} : r_2 < \frac{k_1}{4\pi\hat{D}_2}\text{exp}\left(\frac{-4\pi\Gamma r_3^3}{3}\right)\right\},
\end{equation}
that reproduces the non-linear kinetics of Reaction (\ref{eq:tri_molecular_reaction}) and thus can be used to construct a particle-based simulation of Reaction (\ref{eq:Example_reaction_network}).\par
We will focus on investigating the kinetics that result from the reaction boundary in Equation (\ref{eq:generalised_MM_reaction_condition}) as to our knowledge this constitutes the first example of a proximity-based reaction condition capable of directly reproducing non-linear kinetics. Before proceeding however, it is worth considering what kinds of kinetics are attainable via our method. In the current presentation, our method requires that the original reaction network can be reduced to an equivalent trimolecular reaction in the same form as Reaction (\ref{eq:tri_molecular_reaction}). Treating bimolecular reactions with non-linear reaction rates - such as the original Michaelis-Menten system \cite{michaelis1913kinetik} - is more difficult, since removing the third reactant would also reduce the spatial degrees of freedom that can be utilised when designing the reaction boundary. In principle, the method could be generalised to higher order reactions, i.e. those involving four or more reactants, and since the additional reactants introduce new spatial degrees of freedom it is conceivable that this might allow more exotic non-linear rates to be simulated. An additional restriction can be inferred directly from Equation (\ref{eq:laplace_transform_rate}), namely that the inverse Laplace transform of the reaction rate must exist. Moreover, the resulting function $f$ must be non-negative for all values of $r_3$ since it denotes a distance.


\section{Simulation of non-linear kinetics}


\label{sec:simulation_framework}
To construct a particle-based simulation of trimolecular reactions like Reaction (\ref{eq:tri_molecular_reaction}) we can make use of the methods typically employed by simulations of Smoluchowski's framework. These simulations are often divided into time-driven (TD), and event-driven (ED) approaches. TD approaches progress through time using finite preset timesteps and the state of the system - the position of each molecule - is updated during each timestep. Following the position updates, the distance between relevant reactants can be calculated and reactions are performed according to the imposed reaction condition. In contrast, ED algorithms calculate the first passage times associated with the movement of molecules, enabling accurate sampling of the exact event times. For instance, the distribution for the first passage time for a molecule to leave a section of the domain and the time at which two molecules first approach within a set distance are usually of interest. Event times are sampled from the appropriate distribution and the system is updated according to a time ordered queue of events that is dynamically updated as each event is processed. Inspired by the approach taken by Vijaykumar, Bolhuis and Ten Wolde \cite{hybridSim} we adopt a hybrid methodology that contains both TD and ED components. The simulation switches between these two modes - referred to as TD mode and ED mode henceforth - in order to efficiently track molecule positions and apply reaction conditions. \par
The TD components of our simulation are based on a popular constant timestep algorithm developed by Andrews and Bray in $2004$ \cite{Andrews_2004}. Their algorithm is simple yet accurate, making it easily adaptable to modifications of Smoluchowski's original framework. The original algorithm has been implemented in the Smoldyn software package, which has been used widely in literature \cite{smoldyn_publications_2022}. Smoldyn simulates diffusion and bimolecular reactions with single molecule detail and can be extended in a straightforward manner to higher order reactions \cite{flegg2016smoluchowski}. \par
In TD mode, the simulation progresses through time via a discrete timestep $\Delta t$. The position, $\pos{x}{i}\parentheses{(}{t}{)}$, of each molecule $i$ is updated randomly each step according to \cite{brownianDynamics}
\begin{equation}\label{eq:brownian_dynamics}
    \pos{x}{i}(t+\Delta t) = \pos{x}{i}(t) + \sqrt{2D_i\Delta t}\pos{\xi}{i},
\end{equation}
where $\pos{\xi}{i}$ is a three-dimensional vector of independent, normally distributed, random numbers with unit variance and zero mean. Once the positions have been updated, the separation vectors $\pos{\eta}{2}$ and $\pos{\eta}{3}$ are calculated according to Equation (\ref{eq:etai_def}) and the reaction boundary in Equation (\ref{eq:generalised_MM_reaction_condition}) is tested to determine if any reaction events took place during the last timestep. Equation (\ref{eq:brownian_dynamics}) provides an exact simulation of molecular diffusion, but does not account for the fact that reactants should instantaneously react once their separations satisfy the reaction condition. Molecules do not undergo continuous motion and can `jump' through the reaction boundary, artificially skipping a reaction during the timestep. The issue can be avoided by making the timestep sufficiently small, but this usually requires excessive computational time. Instead, it is common to make use of a numerically derived reaction boundary that is slightly larger than the continuous-time theoretical boundary. The required numerical radius is determined by the desired reaction rate and $\Delta t$ allowing it to be precomputed. Corrections of this form are straightforward for fixed reaction boundaries, but the size of the reaction boundary in Equation (\ref{eq:generalised_MM_reaction_condition}) is a function of $r_3$. As a consequence, we need to compute a series of corrections so that the appropriate value may be retrieved for any value of $r_3$. \par
Reaction (\ref{eq:tri_molecular_reaction}) is trimolecular, but it is convenient to imagine that it is the result of two bimolecular reactions so that the original Smoldyn protocol can be used. We treat molecules of $A$ and $B$ as undergoing a bimolecular reaction to form a complex $AB$ according to the reaction condition
\begin{equation}\label{eq:sim_complex_formation_boundary}
    \Omega_{AB} = \left\{\pos{\eta}{2} : r_2 \le f(0)\right\},   
\end{equation}
where $f$ is given by Equation (\ref{eq:generalised_MM_reaction_boundary}).
Molecules of $AB$ can then react with molecules of $C$ to produce molecules of $P$ so that the system evolves according to
\begin{equation}
\label{eq:sim_reaction}
    \ce{A + B <=> AB} \quad \text{and} \quad \ce{AB + C -> P + C}.
\end{equation}
A molecule of $AB$ is initialised at the centre of diffusion of the reacting $A$ and $B$ molecules,  which is defined by Equation (\ref{eq:x_bari_def}) as
\begin{equation}
    \posbar{x}{2} = \frac{D_1^{-1}}{D_1^{-1}+D_2^{-1}}\pos{x}{1} + \frac{D_2^{-1}}{D_1^{-1}+D_2^{-1}}\pos{x}{2},
\end{equation}
where the labels $1$ and $2$ have been used to refer to the molecule $A$ and $B$ respectively. Similarly, the corresponding separation vector can be calculated according to Equation (\ref{eq:etai_def}),
\begin{equation}
    \pos{\eta}{2} = \pos{x}{2} - \pos{x}{1}.
\end{equation}
Both $\posbar{x}{2}$ and $\pos{\eta}{2}$ undergo independent isotropic linear diffusion with the respective diffusion constants $\bar{D}_2$ (Equation (\ref{eq:x_bar_diff_coeff})) and $\hat{D}_2$ (Equation (\ref{eq:eta_i_diff_coeff})) \cite{flegg2016smoluchowski}, and can be updated in TD mode using Equation (\ref{eq:brownian_dynamics}).\par
\begin{figure}
\centering
\begin{subfigure}[b]{0.75\textwidth}
   \includegraphics[width=1\linewidth]{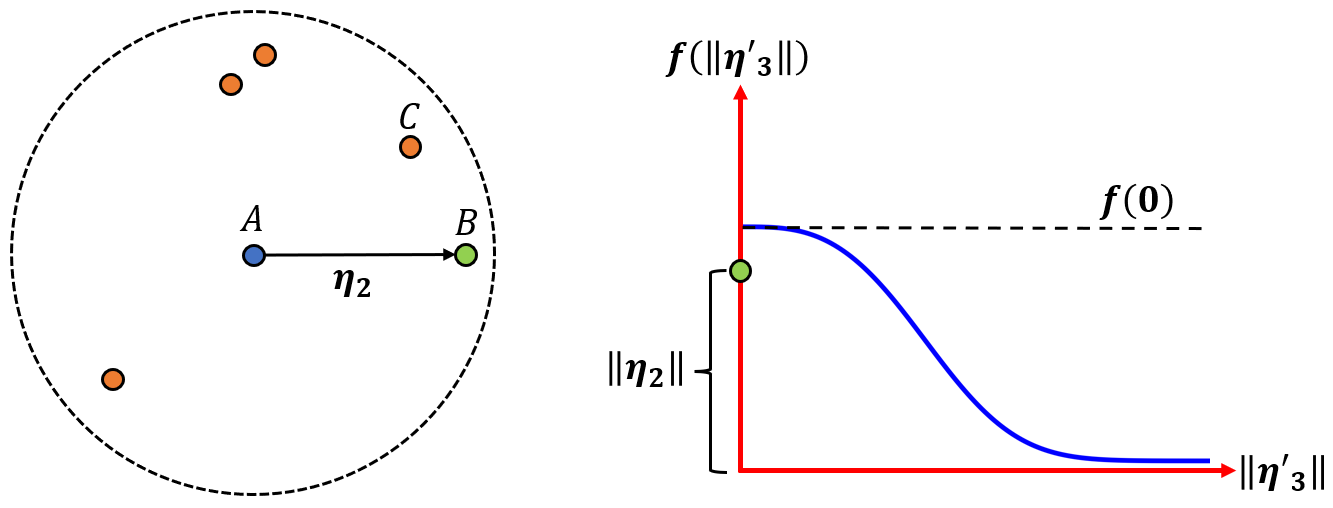}
   \caption{}
   \label{fig:Ng1} 
\end{subfigure}
\begin{subfigure}[b]{0.75\textwidth}
   \includegraphics[width=1\linewidth]{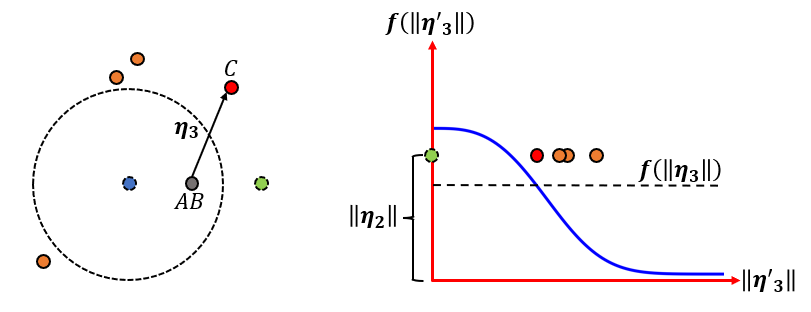}
   \caption{}
   \label{fig:Ng2}
\end{subfigure}
\caption{(a) A pair of molecules $A$ and $B$ - depicted as a blue and green point respectively - are simulated as undergoing a bimolecular reaction to form a molecule of $AB$ whenever their separation $\pos{\eta}{2}$ is such that $||\pos{\eta}{2}|| \le f(0)$. This reaction condition (see Equation (\ref{eq:sim_complex_formation_boundary})) defines a spherical boundary of radius $f(0)$ centred on $A$, which is shown as a dotted circle. The radius of the boundary represents the greatest separation between the two molecules at which it is possible for Reaction (\ref{eq:tri_molecular_reaction}) to occur according to the reaction boundary in Equation (\ref{eq:generalised_MM_reaction_condition}) and is illustrated in the graph on the right as a dotted line. (b) When $A$ and $B$ react, a molecule of $AB$ (the grey point) is initialised at their centre of diffusion, $\posbar{x}{2}$. The separation, $\pos{\eta}{3}$, between $AB$ and the closest molecule of $C$ (the red point) defines the radius of the reaction boundary that surrounds $A$ as shown by the dotted circle on the left and by the dotted line on the graph of the reaction boundary from Equation (\ref{eq:generalised_MM_reaction_condition}) on the right. $\posbar{x}{2}$ and $\pos{\eta}{2}$ diffuse independently and can be updated without explicitly tracking $A$ and $B$, although the positions of the individual molecules can be recovered at any time as indicated by the dotted outlines of the blue and green points. If $||\pos{\eta}{2}|| \le f(||\pos{\eta}{3}||)$ during any timestep a reaction is performed. Similarly, if $||\pos{\eta}{2}|| > f(0)$ the molecule of $AB$ is dissolved and $A$ and $B$ are reintroduced at their respective positions.}
\label{fig:X_formation}
\end{figure}

Reaction (\ref{eq:sim_reaction}) is deceptively similar to Reaction (\ref{eq:Example_reaction_network}) with the complexes $AB$ and $X$ appearing interchangeable. However, $AB$ is not a chemical species. Instead a molecule of $AB$ is constructed within the simulation purely as a convenient way to track any pair of $A$ and $B$ molecules that are close enough to react with a molecule of $C$. Pairs of $A$ and $B$ molecules that are not close enough to form one of these fictitious complexes are treated as being nonreactive with $C$, which avoids unnecessary testing of the reaction condition. If the constituents of a molecule of $AB$ move far enough apart that reaction with $C$ becomes impossible the complex is dissolved and $A$ and $B$ molecules are reintroduced at the positions
\begin{equation}
    \pos{x}{1} = \posbar{x}{2} - \frac{D_2^{-1}}{D_1^{-1} + D_2^{-1}}\pos{\eta}{2} \quad \text{and}\quad \pos{x}{2} = \posbar{x}{2} + \frac{D_1^{-1}}{D_1^{-1} + D_2^{-1}}\pos{\eta}{2},
\end{equation}
respectively. When viewed in this way, Reaction (\ref{eq:tri_molecular_reaction}) can be thought of as a bimolecular reaction between $A$ and $B$ where the reaction boundary for each pair of molecules is determined by the proximity of the associated $AB$ complex to the closest molecule of $C$. The position of the complex is the centre of diffusion of the $A$ and $B$ molecules allowing for simple calculation of the separation to each molecule of $C$ using Equation (\ref{eq:etai_def}),
\begin{equation}
    \pos{\eta}{3} = \pos{x}{3} - \posbar{x}{2},
\end{equation}
where the subscript $3$ is associated with a particular molecule of $C$. Once the minimum value of $\pos{\eta}{3}$ has been found, the reaction boundary in Equation (\ref{eq:generalised_MM_reaction_condition}) can be checked to determine if $A$ and $B$ are close enough for a reaction to occur, as shown in Fig.~\ref{fig:X_formation}. The relevant reaction radius is selected from a continuum of radii that need to be corrected to account for the use of finite timesteps. To enable this, we precompute a table of corrections following the protocol described in \cite{Andrews_2004} so that the numerical reaction radius that corresponds to any value of $f(r_3)$ can be retrieved efficiently.\par

\begin{figure}
\centering
\includegraphics[width=1\linewidth]{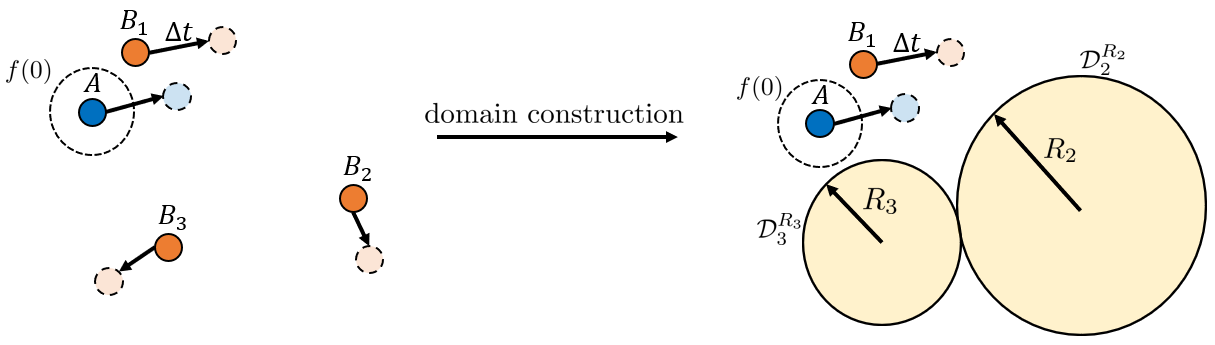}
\caption{The molecule labelled $B_1$ is close to the reactive molecule $A$ and its position must be tracked explicitly and updated each timestep $\Delta t$ to enable the reaction condition - shown as a dotted circle - to be checked. However, the molecules $B_2$ and $B_3$ are too far from $A$ to react within the next timestep and their positions do not need to be known exactly until $A$ approaches close enough that the reaction condition needs to be tested. To avoid computing position updates for $B_2$ and $B_3$ they are placed in single particle protective domains, labelled $\mathcal{D}^{R_2}_2$ and $\mathcal{D}^{R_3}_3$ respectively, until they either escape, or $A$ approaches close enough to the domain to burst it.}
\label{fig:protective_domain_construction}
\end{figure}

Through the use of $AB$ molecules we are able to avoid unnecessary testing of the reaction condition, but a significant amount of time can still be wasted propagating $A$ and $B$ molecules that are not close enough to be considered reactive. This is a common criticism of TD methods and the issue is mitigated by switching the simulation of such molecules to ED mode. The exact position of a molecule only needs to be known if it could be involved in a reaction in the next timestep. Molecules that are isolated from other relevant reactants do not need to be tracked explicitly, and instead can be placed in single particle domains equivalent to those defined in eGFRD \cite{egfrd1,egfrd2}. The introduction of protective domains means that at any time the simulation may contain a mixture of molecules in TD and ED mode, as shown in Fig.~\ref{fig:protective_domain_construction}. The escape time of each ED molecule is placed in a queue, which is used to determine if an escape event will occur within the next TD timestep. If an event is scheduled to occur between $\parentheses{(}{t,t+\Delta t}{]}$ the corresponding event time, $t_\text{event}$, is removed from the queue. The event is then processed, and the position of all molecules in TD mode are updated using the timestep, $\Delta t^\prime = t_\text{event}-t$. Since a molecule in TD mode may approach a protective domain before the domain's escape time, it is possible for a reaction to occur between the TD molecule and the molecule within the domain. We cannot be sure of the exact position of the molecule within the protective domain, so to avoid missing a potential reaction we must prematurely dissolve, or burst the domain. Upon bursting a domain, the position of the enclosed molecule is sampled as described in \cite{egfrd2}. If the newly sampled position is too close to a neighbouring domain this domain is also burst and this process continues until all domains have been burst or are sufficiently isolated from any of the molecules in TD mode. Any molecule that has escaped its domain or had it burst, is placed in TD mode until it becomes sufficiently isolated from the other reactants to be placed in a new protective domain.
\par
Protective domains allow the simulation to make large adaptive jumps forward in time during uninteresting periods where all the reactants are too far from each other for reaction events to be possible. If two or more reactants are close enough that reaction conditions need to be checked, then only the relevant molecules need be switched to TD mode, greatly reducing the number of reactant combinations that have to be considered. Protective domains can be applied to the fictitious $AB$ molecules, but they are more like the pair domains used in eGFRD than the single particle domains described thus far. Similar to a molecule of $A$, $B$, or $C$, an $AB$ molecule might escape its protective domain by simply reaching its boundary, or the domain might be burst because a $C$ molecule moved close enough that a reaction is possible. However, an additional domain needs to be constructed for $\pos{\eta}{2}$ since the complex should be dissolved if $r_2 > f(0)$ and a reaction should occur if the closest $C$ molecule is such that $r_2 \le f(r_3)$. This means that each molecule of $AB$ has two protective domains from which it can escape; one for each of $\posbar{x}{2}$ and $\pos{\eta}{2}$. The protective domain for $\posbar{x}{2}$ is equivalent to the single particle domain, and the problem is identical to that of the `centre of motion' considered in eGFRD. Similarly, the problem for $\pos{\eta}{2}$ is analogous to that of the `inter-particle vector' in eGFRD, but differs crucially in that the inner boundary condition that corresponds to the reaction of the $A$ and $B$ molecules is no longer static. Instead, this boundary is determined by the proximity of the nearest molecule of $C$ preventing reuse of the standard pair domain methods. If we ignore this inner boundary, it is possible to find the greens function for an $\pos{\eta}{2}$ domain that only has an outer boundary at $r_2 = f(0)$. This additional domain means that both $\posbar{x}{2}$ and $\pos{\eta}{2}$ have to be sampled when an escape event occurs or if the domain is burst, but otherwise the domain functions similarly to a single particle domain. Unfortunately, the utility of such domains is diminished by the fact that reactions can occur even when $r_3$ is large so long as $r_2$ is sufficiently small and due to this we did not implement protective domains for $AB$ molecules during the simulation.

\section{\label{sec:numerical_results}Numerical results}

In this section, we present the results of a series of particle-based simulations of Reaction (\ref{eq:tri_molecular_reaction}) conducted using an implementation of the simulation framework discussed in Section~\ref{sec:simulation_framework}. The first test considers a well-mixed trimolecular system and is designed to validate our implementation. While the second test explores whether the reaction boundary in Equation (\ref{eq:generalised_MM_reaction_condition}) reproduces the non-linear behaviour described in Equation (\ref{eq:MM_degradation_rate_a}) as $c$ is varied.
\subsection{\label{subsec:trimolecular_test}Well-mixed trimolecular test system}
To verify our simulation method we consider the trimolecular system
\begin{equation}\label{eq:first_test_reaction}
    \ce{A + B + C ->[k_3\parentheses{(}{c}{)}] C}, \quad \ce{$\emptyset$ ->[k_0] A},
\end{equation}
where $k_0$ is a zeroth order rate constant that controls the production of $A$ and $k_3\parentheses{(}{c}{)}$ is given in Equation (\ref{eq:trimolecular_rate_constant}).
The population of $A$ molecules in the system is governed by the Law of Mass Action and when expressed in terms of molecular populations rather than concentrations may be written
\begin{equation}\label{eq:numerical_testSystem}
    \frac{dN_A}{dt} = k_0^\prime - \frac{k_1^\prime }{\Gamma^\prime + N_C}N_AN_BN_C,
\end{equation}
where $N_A$, $N_B$ and $N_C$ are the respective number of $A$, $B$ and $C$ molecules at time $t$, and we have defined the parameters
\begin{equation}
    k_0^\prime = Vk_0, \quad k_1^\prime = \frac{k_1}{V}, \quad \text{and} \quad \Gamma^\prime = V\Gamma.
\end{equation}
The simulation is conducted in a well-mixed domain that is a dimensionless unit cube with periodic boundary conditions. Within the domain we randomly place a single immortal molecule of $B$, so that $N_B = 1$ for the duration of the simulation. Accounting for this single molecule of $B$ and defining the dimensionless variables
\begin{equation}
    \bar{N}_A = \frac{k_1^\prime N_B}{k_0^\prime}N_A, \quad \bar{N}_c = \frac{N_C}{\Gamma^\prime}, \quad \text{and} \quad \tau = k_1^\prime N_B t,
\end{equation}
Equation (\ref{eq:numerical_testSystem}) becomes
\begin{equation}\label{eq:numerical_testSystem_dimensionless}
    \frac{d\bar{N}_A}{d\tau} = 1 - \frac{ \bar{N}_A\bar{N}_C}{1 + \bar{N}_C}.
\end{equation}
\par
We set $\bar{N}_C = 5$ by choosing $\Gamma^\prime = 1$ and placing $N_C = 5$ immortal $C$ molecules uniformly at random within the domain. Using dimensionless diffusion coefficients $D_1 = D_2 = D_3 = 1$ and time steps of $\Delta \tau = 2\times 10^{-6}$ we simulate the system for a non-dimensional duration of $\tau = 10$. The steady-state distribution of $\bar{N}_A$ is governed by the master equation associated with this birth-death process, from which we expect a Poisson distribution \cite{erban2007practical},
\begin{equation}
\label{eq:NAbar_distribution}
    \bar{N}_A = \text{Pois}\parentheses{(}{\frac{1+\bar{N}_C}{\bar{N}_C}}{)}.
\end{equation}
We choose the scaled rate constants $k_0^\prime = 1$ and $k_1^\prime = 0.2$ such that $\bar{N}_A = 0.2 N_A$ and at the beginning of each simulation, we sample $N_A$ from the expected steady-state population, $N_A = \text{Pois}\parentheses{(}{6}{)}$, and uniformly distribute the molecules throughout the volume. The simulation was repeated $3\times 10^4$ times and the population of $A$ molecules was sampled at the conclusion of each simulation. \par
\begin{figure}
    \centering
    \includegraphics[width=\linewidth]{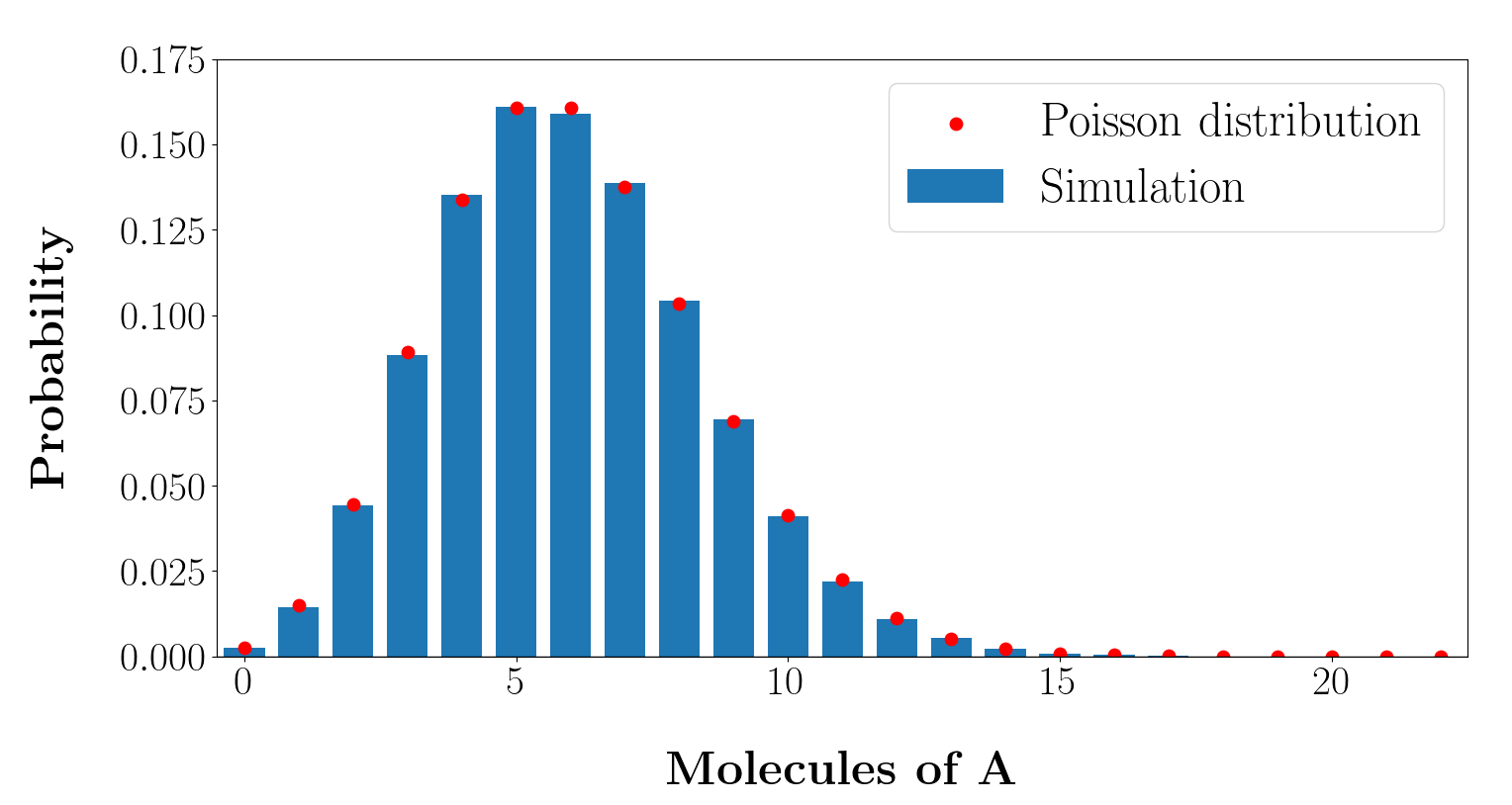}
    \caption{The distribution of the number of molecules of $A$ ($N_A$) present in System (\ref{eq:first_test_reaction}) obtained from an implementation of the particle-based simulation described in Section~\ref{sec:simulation_framework}. The value of $N_A$ in each simulation is sampled after a dimensionless time of $\tau = 10$ has been reached to ensure the system has reached steady state. The blue bars depict the distribution sampled from $3\times 10^4$ simulations of System \ref{eq:first_test_reaction}, while the red dots represent the Poisson distribution predicted by the Law of Mass Action. The mean, $6.00\pm0.01$, and the variance, $5.98 \pm 0.05$, of the simulated distribution does not deviate significantly from that of the theoretical distribution, which has a mean and variance of $6$. The maximum error associated with the histogram is less than $10^{-4}$ and would not be visible in this figure.}
    \label{fig:trimolecular_test}
\end{figure}

 
In Fig.~\ref{fig:trimolecular_test}, the sampled distribution of $N_A$ (blue bars) is compared with the expected Poisson distribution (red dots). Neither the mean, $6.00 \pm 0.01$, nor the variance, $5.98 \pm 0.05$, of the simulated distribution deviate significantly from the initial Poisson distribution. This indicates that the kinetics observed in our particle-based simulation closely resemble the theoretical kinetics for the choices of reaction parameters considered. However, if the reaction parameters are not chosen carefully, one may expect to observe some errors arising from the fact that the boundary in Equation (\ref{eq:generalised_MM_reaction_condition}) only approximates the actual reaction boundary required to reproduce the kinetics of Reaction (\ref{eq:tri_molecular_reaction}). Moreover, the reaction boundary is further altered during the simulation in an attempt to correct for the errors introduced by the use of finite timesteps.

\subsection{Non-linear kinetics}
System (\ref{eq:first_test_reaction}) degrades $A$ at a rate that grows non-linearly in $N_C$ in accordance with Equation (\ref{eq:numerical_testSystem}). The associated dimensionless reaction rate is given by the inverse of the mean of the steady-state distribution for $\bar{N}_A$ in Equation (\ref{eq:NAbar_distribution}),
\begin{equation}
\label{eq:expected_non_linear_rate}
    \bar{K} = \frac{\bar{N}_C}{1+\bar{N}_C}.
\end{equation}
To demonstrate that our framework reproduces this expected non-linear behaviour, we use the same methodology described in \ref{subsec:trimolecular_test}, to simulate System (\ref{eq:first_test_reaction}) with $\Gamma = 2$, and $k_1 = 0.3$, while $k_0$ and $V$ are chosen so that $\bar{N}_A = 0.3N_A$. These reaction parameters were selected since they yield convenient dimensional reaction rates for several of the $\bar{N}_C$ values considered. Although, any set that did not violate the assumptions used to derive the reaction boundary in Equation (\ref{eq:generalised_MM_reaction_condition}) - namely that the boundary is thin in $r_2$ and long in $r_3$ - could be considered. \par
\begin{figure}
    \centering
    \includegraphics[width=\linewidth]{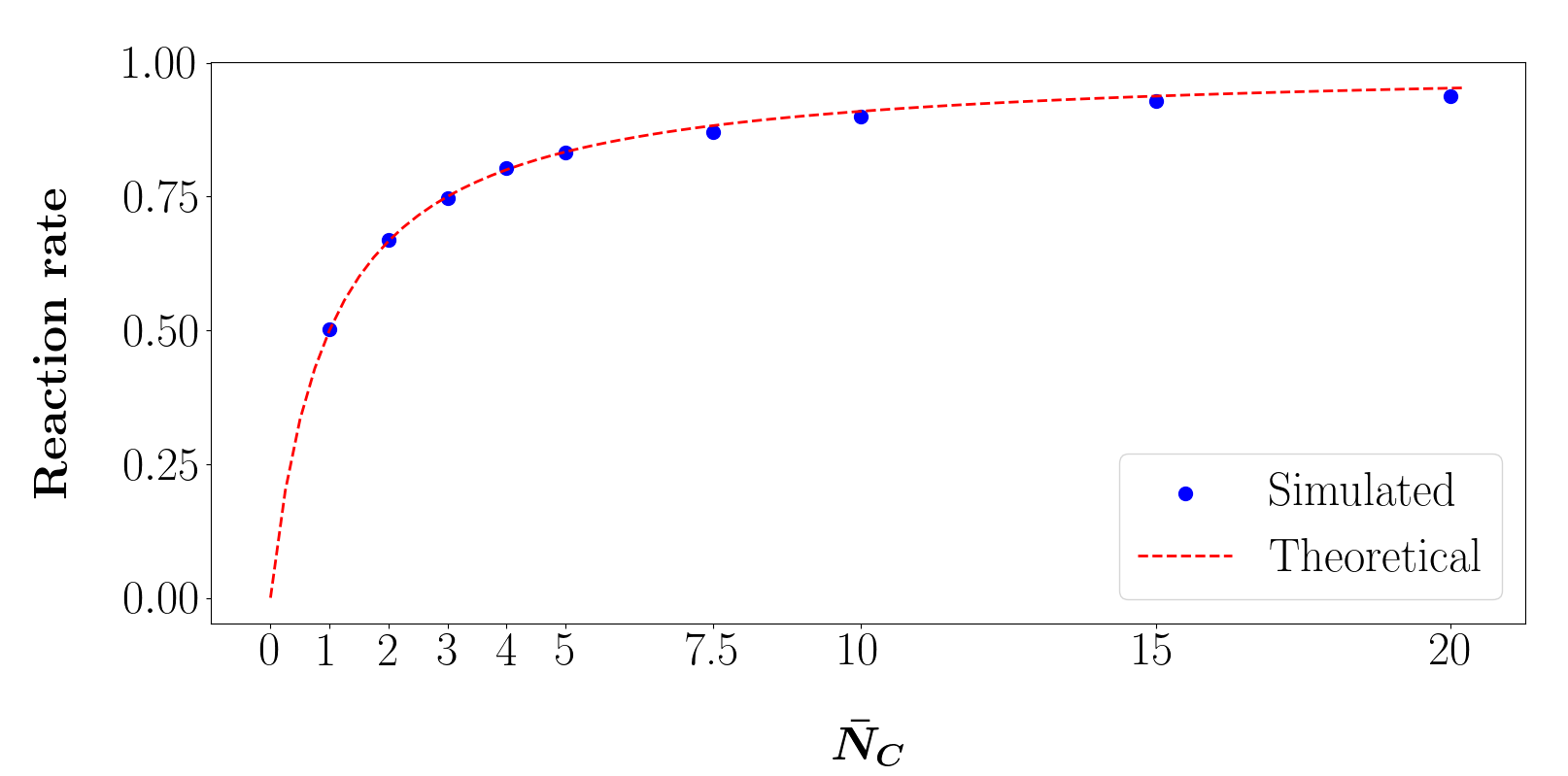}
    \caption{The dimensionless reaction rate, $\bar{K}$, obtained from particle-based simulations of System (\ref{eq:first_test_reaction}) plotted as a function of $\bar{N}_C$. The theoretical reaction rate given in Equation (\ref{eq:expected_non_linear_rate}) is plotted as a red dashed line and compared to the simulated rate shown by the blue points for $\bar{N}_C \in \left\{1,2,3,4,5,7.5,10,15,20\right\}$. The shape of the reaction boundary (see Equation (\ref{eq:generalised_MM_reaction_condition})) is kept constant by setting $\Gamma = 2$ and $k_1 = 0.3$ for all simulations. For $\bar{N}_C \geq 5$ we set $k_0 = 1$ and conduct the simulation in a unit cube, while for $\bar{N}_C < 5$ the volume is increased (and $k_0$ adjusted accordingly) to ensure that the simulation contains an adequate number of independent $C$ molecules. The simulated reaction rate was calculated from $1\times 10^4$ simulations for each value of $\bar{N}_C$ and all resulting errors are less than $5\times 10^{-3}$ so the associated error bars would not be visible in this figure. }
    \label{fig:nonlinear_test}
\end{figure}

 
By altering the value of $N_C$ accordingly, we perform $1\times 10^4$ simulations for each value in \\$\bar{N}_C \in \left\{1,2,3,4,5,7.5,10,15,20\right\}$ and calculate the corresponding dimensionless reaction rates. The simulated reaction rate is calculated by taking the inverse of the mean of the steady-state distribution obtained for $\bar{N}_A$ for each value of $\bar{N}_C$. These simulated rates are shown by the blue points in Fig.~\ref{fig:nonlinear_test}, where they are compared against a plot of Equation (\ref{eq:expected_non_linear_rate}) shown by the red dashed line. The simulated reaction rate agrees with the theoretical rate for all the values of $\bar{N}_C$ considered, and can be seen to reproduce the characteristic non-linear behaviour predicted by Equation (\ref{eq:expected_non_linear_rate}). However, there is some indication that the simulated reaction rate has a lower horizontal asymptote than predicted, and it is possible that the rates would eventually diverge if $\bar{N}_C$ was increased further. This is likely a result of the fact that the chance a reaction event is missed during the simulation increases as the number of $C$ molecules is increased due to crowding. In addition, it should be noted that initially we considered $k_0 = 1$ and $V=1$ for all of our simulations, but for these parameters we found that the simulated rate significantly exceeded the theoretical rate for $\bar{N}_C < 5$. We attributed the majority of this discrepancy to the fact that the simulated system becomes an increasingly poor reproduction of the situation considered in Section \ref{sec:generalised_reaction_conditions}.\par
\begin{figure}
    \centering
    \includegraphics[width=\linewidth]{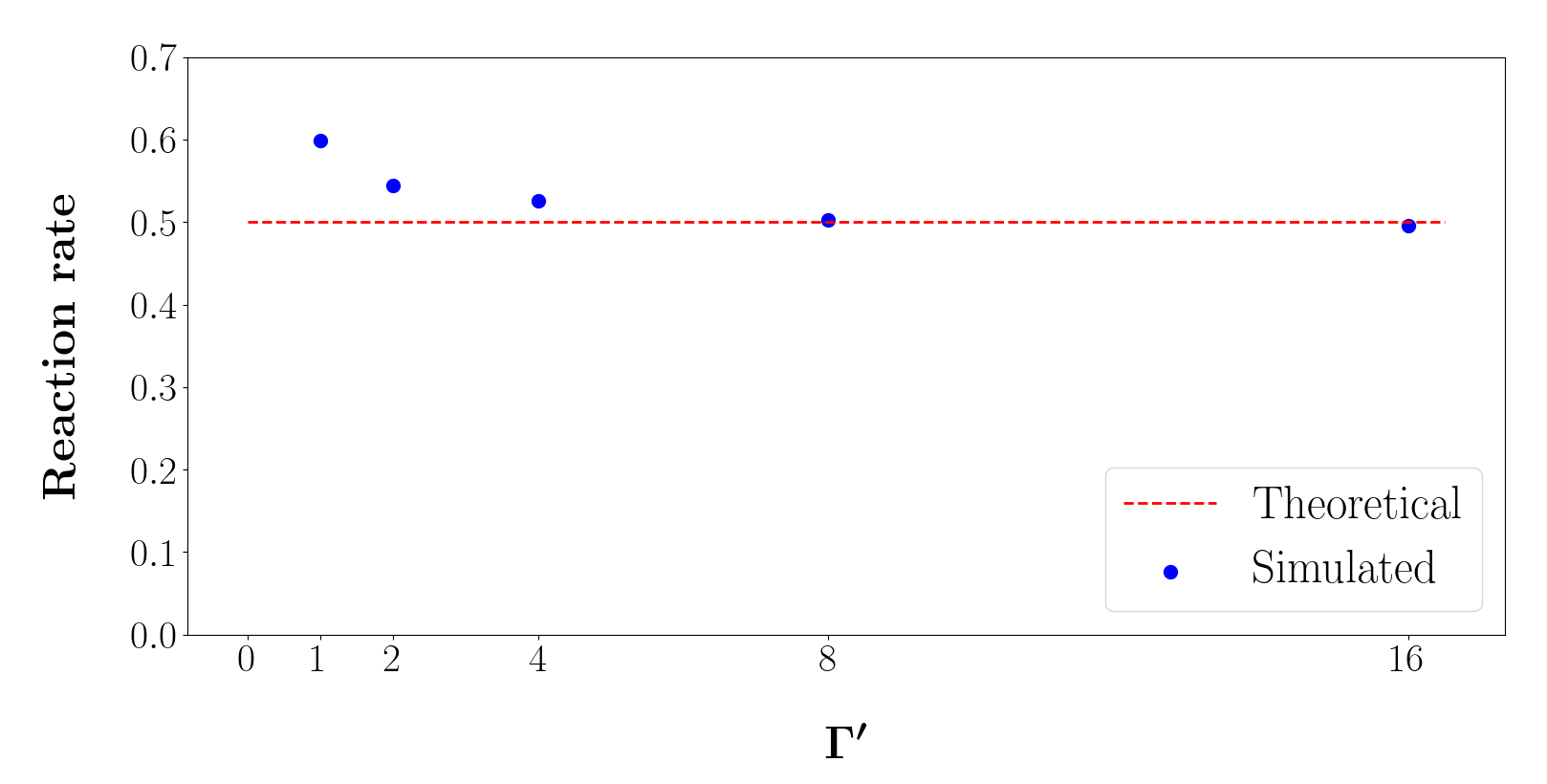}
    \caption{The dimensionless reaction rate, $\bar{K}$, obtained from particle-based simulations of System (\ref{eq:first_test_reaction}) with $\bar{N}_C = 1$ plotted for $\Gamma^\prime \in \left\{1,2,4,8,16\right\}$. The theoretical reaction rate given in Equation (\ref{eq:expected_non_linear_rate}) is plotted as a red dashed line and compared to the simulated reaction rate shown by the blue points. The shape of the reaction boundary (see Equation (\ref{eq:generalised_MM_reaction_condition})) is kept constant by setting $\Gamma = 2$ and $k_1 = 0.3$, and, $\Gamma^\prime = V\Gamma$, is instead altered by increasing or decreasing the simulation volume $V$. The value of $k_0$ is adjusted accordingly such that the ratio, $k_1/\left(V^2k_0\right)$, is $0.3$ for each simulation and the number of independent molecules of $C$ is set to $N_C = \Gamma^\prime$, so that $\bar{N}_C = N_C/\Gamma^\prime = 1$. We can see that as $\Gamma^\prime$ - and hence $N_C$ - is decreased, the simulated reaction rate increases and exceeds the theoretical reaction rate for $\Gamma^\prime \leq 4$ ($N_C \leq 4$). There is good agreement between the simulated and theoretical reaction rates for $\Gamma^\prime = 8$ and $\Gamma^\prime = 16$, indicating that the simulation contains sufficiently many independent molecules of $C$ to be a good approximation of the effectively infinite population considered in Section \ref{sec:generalised_reaction_conditions}. The simulated reaction rate was calculated from $5\times 10^3$ simulations for each value of $\Gamma^\prime$ and all resulting errors are less than $4\times 10^{-3}$ so the associated error bars would not be visible in this figure.}
    \label{fig:gamma_test}
\end{figure}

 
The derivation of the reaction boundary in Equation (\ref{eq:generalised_MM_reaction_condition}) is only valid in the limit that $V \to \infty$ so that an effectively infinite number of independent $C$ molecules are present regardless of the concentration of $C$. In the simulation we approximate this infinite population by placing periodic boundary conditions on our volume, however the `image' position of any molecule within the volume is perfectly correlated with the molecule's actual position. Therefore, if for example $N_C = 2$, then the simulation would only contain two independent molecules of $C$. This can be corrected by increasing $N_C$, but necessitates an increase in $\Gamma^\prime$ if $\bar{N}_C$ is to be kept constant. Since $\Gamma$ controls the shape of the reaction boundary, we leave it unchanged at $\Gamma = 2$ and increase $\Gamma^\prime$ by expanding the simulation volume instead. Similarly, we retain $k_1 = 0.3$ and alter $k_0$ - which does not impact the shape of the reaction boundary - so that $\bar{N}_A = 0.3N_A$ for all values of $\bar{N}_C$ considered. In this way we were able to obtain the simulated reaction rates for $\bar{N}_C \in \left\{1,2,3,4\right\}$ that are shown in Fig.~\ref{fig:nonlinear_test}. Fig.~\ref{fig:gamma_test} demonstrates the corrections obtained via this method for $\bar{N}_C = 1$. Here we plot the simulated reaction rate as a function of $\Gamma^\prime$, which is altered by changing $V$ and keeping $\Gamma$ fixed. We then set $N_C = \Gamma^\prime$ so that $\bar{N}_C$ remains $1$. The value of $\bar{K}$ is calculated from $5\times 10^3$ simulations for each value in $\Gamma^\prime \in \left\{1,2,4,8,16\right\}$ and is shown in blue, while the theoretical rate given by Equation (\ref{eq:expected_non_linear_rate}) is shown by the red dotted line. We can see that as $\Gamma^\prime$ - and hence the number of independent $C$ molecules - decreases, the simulated reaction rate increases and exceeds the theoretical reaction rate for $\Gamma^\prime \leq 4$ ($N_C \leq 4$). For $\Gamma^\prime = 8$ and $\Gamma^\prime = 16$ the simulated reaction rate agrees with the theoretical rate, indicating that the simulation contains sufficiently many independent molecules of $C$ to be a good approximation of the effectively infinite population considered in Section \ref{sec:generalised_reaction_conditions}. Finally, we note that the simulated reaction rate does appear to decrease slightly when $\Gamma^\prime$ - and hence $N_C$ - is increased from $8$ to $16$ and this is likely a consequence of the crowding observed earlier.

\section{\label{sec:conclusions}Conclusions}
We proposed a modification to Smoluchowski's model of reaction diffusion systems that enables us to reproduce non-linear reaction rates characteristic of enzyme kinetics. While Smoluchowski's bimolecular reaction condition is unable to correctly incorporate the fast reactions associated with the formation of enzyme-substrate complexes, the reaction boundary in Equation (\ref{eq:generalised_MM_reaction_condition}) allows us to reproduce the non-linear kinetics of Reaction (\ref{eq:Example_reaction_network}) without needing to explicitly simulate these reactions. Our reaction condition differs from the static bimolecular conditions traditionally used in derivatives of Smoluchowski's framework as the size of the boundary is determined by the relative proximity of the reactants. Although Reaction (\ref{eq:Example_reaction_network}) is trimolecular, we have shown that it can be thought of as a bimolecular reaction between $A$ and $B$ where the size of the reaction boundary is determined by the proximity of the third molecule $C$. This view allows trimolecular reactions of this form to be easily and efficiently incorporated into time driven simulations of Smoluchowski's framework. In addition, we have identified several components of eGFRD that can be easily implemented in the presence of our generalised reaction boundaries. Leveraging these ideas we have conducted proof of concept simulations that demonstrate our reaction boundary reproduces the expected non-linear kinetics. While the theory presented here is limited to systems that can be reduced to a single trimolecular reaction  similar in form to Reaction (\ref{eq:tri_molecular_reaction}), in a future publication we hope to extend our framework to a wider variety of enzymatic systems.


\appendix
\setcounter{equation}{0}
\renewcommand\theequation{A.\arabic{equation}}

\section{Evolution of the closest molecule}\label{sec:AppendixA}
The scaled probability density function, $\PhiFunc$, defined originally in Equation (\ref{eq:Phi_definition}),
\begin{equation}
    \PhiFunc = \frac{\probGiven[]{\pos{\eta}{3}}{t}{\pos{\Eta}{3}}}{c},
\end{equation}
is proportional to the probability that a particular molecule of $C$ is associated with $\pos{\eta}{3}$ at time $t$, given that we know it is the closest $C$ to the origin. We assume our system is large and consider the limit, $N_C \to \infty$, so that $\PhiFunc$ is given by Equation (\ref{eq:Phi_definition_in_limit}),
\begin{equation}
    \PhiFunc = \phiFunc \text{exp}\left(-c\int_{V_3}\phiFuncPrime dV_3'\right),
\end{equation}
where $\phiFunc$ is the scaled probability that the molecule of $C$ is associated with $\pos{\eta}{3}$ originally defined in Equation (\ref{eq:phi_definition}),
\begin{equation}
    \phiFunc = \frac{N_C\prob[]{\pos{\eta}{3}}{t}}{c}.
\end{equation}
\par
To derive the governing equation for $\PhiFunc$ we consider $\mathcal{L}_3\PhiFunc$ where
\begin{equation}
    \mathcal{L}_3 \equiv \frac{\partial}{\partial t} - \hat{D}_3\hat{\nabla}^2_3,
\end{equation}
is the diffusion operator on the $3$-dimensional space spanned by $\pos{\eta}{3}$, originally defined in Equation (\ref{eq:diffusion_eta3}). The time derivative of $\PhiFunc$ is given by
\begin{equation}\label{eq:time_derivative}
    \Phi_t = g\phi_t - c\phi g\int_{V_3}\phi_t\left(\pos{\eta'}{3},t\right) dV_3',
\end{equation}
where we have used the $t$ subscript to denote differentiation with respect to time and defined 
\begin{equation}
    g \equiv \gFunc = \text{exp}\left(-c\int_{V_3}\phiFuncPrime dV_3'\right),
\end{equation}
for notational convenience. We can also take Laplacian of $\Phi$ with respect to the coordinates of $\pos{\eta}{3}$
\begin{equation}
    \hat{\nabla}^2_3 \Phi = g\hat{\nabla}^2_3\phi + 2\left(\hat{\nabla}_3\phi\right)\cdot\left(\hat{\nabla}_3g\right) + \phi \hat{\nabla}^2_3g,
\end{equation}
which when combined with Equation (\ref{eq:time_derivative}) yields
\begin{equation}\label{eq:L3_on_Phi}
    \mathcal{L}_3 \Phi = g\phi_t - c\phi g\int_{V_3}\phi_t\left(\pos{\eta'}{3},t\right) dV_3'- \hat{D}_3\left[g\hat{\nabla}^2_3\phi - 2\left(\hat{\nabla}_3\phi\right)\cdot\left(\hat{\nabla}_3g\right) - \phi \hat{\nabla}^2_3g\right].
\end{equation}
\par
Recalling that $\mathcal{L}_3\phi = 0$ from Equation (\ref{eq:diffusion_eta3}) and applying the Divergence theorem we find
\begin{equation}\label{eq:L3_on_Phi_simplified1}
    \begin{split}
    \mathcal{L}_3 \Phi &= \hat{D}_3\left[-c\phi g\int_{V_3}\hat{\nabla}^2_3\phi\left(\pos{\eta'}{3},t\right) dV_3' - 2\left(\hat{\nabla}_3\phi\right)\cdot\left(\hat{\nabla}_3g\right) - \phi \hat{\nabla}^2_3g\right]\\
    &= - c\hat{D}_3\phi g\oint_{S_3}\left(\hat{\nabla}_3\phi\left(\pos{\eta'}{3},t\right)\right)\cdot \hat{\pos{r}{3}} dA_3' - 2\hat{D}_3\left(\hat{\nabla}_3\phi\right)\cdot\left(\hat{\nabla}_3g\right) - \hat{D}_3\phi \hat{\nabla}^2_3g,
\end{split}
\end{equation}
where $S_3$ is the surface of a sphere of radius $r_3 = ||\pos{\eta}{3}||$, $dA_3'$ is an elemental area on that surface and $\hat{\pos{r}{3}}$ is the unit outward facing normal vector. The diffusion in the $\pos{\eta}{3}$ coordinates is isotropic and so $\phi$ is independent of orientation. That is, $\phi$ only has radial dependence, so we have
\begin{equation}\label{eq:nabla_g}
\begin{split}
    \hat{\nabla}_3 g &= \hat{\nabla}_3 \text{exp}\left(-c\int_{V_3}\phiFuncPrime dV_3'\right) \\&= -cg\phi\left(\oint_{S_3} dA_3'\right)\hat{\pos{r}{3}}.
\end{split}
\end{equation}
By the same reasoning we are able to take the integrand outside of the integral in Equation (\ref{eq:L3_on_Phi_simplified1}) and by substituting in Equation (\ref{eq:nabla_g}) we obtain
\begin{equation}
    \begin{split}
    \mathcal{L}_3 \Phi &= -\hat{D}_3 \parentheses{[}{\parentheses{(}{\hat{\nabla}_3 \phi}{)}\cdot \parentheses{(}{\hat{\nabla}_3g}{)} + \phi \hat{\nabla}^2_3g}{]}\\
    &= -\hat{D}_3\hat{\nabla}_3\cdot \parentheses{(}{\phi\hat{\nabla}_3g}{)}\\
    &= \hat{D}_3\hat{\nabla}_3 \cdot \parentheses{(}{cg\phi^2\left(\oint_{S_3} dA_3'\right)\hat{\pos{r}{3}}}{)}\\
    &= \hat{D}_3\hat{\nabla}_3 \cdot \parentheses{(}{4\pi r_3^2 c\phi\Phi\hat{\pos{r}{3}}}{)}.
\end{split}
\end{equation}
That is,
\begin{equation}
    \frac{\partial \PhiFunc}{\partial t} = \hat{D}_3\hat{\nabla}^2_3\PhiFunc + \hat{D}_3\hat{\nabla}_3 \cdot \parentheses{(}{4\pi r_3^2 c\phi\PhiFunc\hat{\pos{r}{3}}}{)},
\end{equation}
as stated in Equation (\ref{eq:Phi_evolution}).

\bibliographystyle{siamplain}
\bibliography{references}
\end{document}